\documentclass[
reprint,
superscriptaddress,
amsmath,
amssymb,
pra,
floatfix
]{revtex4-2}

\usepackage[T1]{fontenc}
\usepackage{mathastext}
\usepackage{xcolor}
\usepackage{braket}

\usepackage{graphicx}
\usepackage{dcolumn}
\usepackage{bm}
\usepackage{array}
\usepackage{url}
\usepackage{hyperref}
\usepackage[normalem]{ulem} 

\begin{document}

\preprint{APS/123-QED}

\title{Communication Bandwidth Limits of Rydberg-Atom Receivers}
%

\author{Rajavardhan~Talashila}
\affiliation{National Institute of Standards and Technology, Boulder, Colorado 80305, USA}
\affiliation{Department of ECEN, University of Colorado, Boulder, Colorado 80309, USA}
\author{Stone~B.~Oliver}
\affiliation{National Institute of Standards and Technology, Boulder, Colorado 80305, USA}
\affiliation{Department of Physics, University of Colorado, Boulder, Colorado 80309, USA}
\author{Noah~Schlossberger}
\affiliation{National Institute of Standards and Technology, Boulder, Colorado 80305, USA}
\author{Christopher~L.~Holloway}
\affiliation{National Institute of Standards and Technology, Boulder, Colorado 80305, USA}
\email[]{christopher.holloway@nist.gov}
\author{Nikunjkumar~Prajapati} 
\affiliation{National Institute of Standards and Technology, Boulder, Colorado 80305, USA}

\date{\today}

\begin{abstract}
Increasing communication bandwidth increases the integrated noise power, while nonlinear distortion limits the signal power that can be used to compensate for this increase. In Rydberg-atom receivers, electromagnetically induced transparency and Autler–Townes spectra define a nonlinear transfer function that generates intermodulation products. Here we parameterize this response using a single  spectroscopic parameter, the standard deviation $\sigma$ of the Gaussian fit to the transparency lineshape. The model predicts fundamental and higher-order intermodulation products across signal and local-oscillator amplitudes and yields analytical conditions for selective intermodulation cancellation that agree with experiment. It also reproduces the measured error vector magnitude response. At a given signal power, intermodulation distortion and  photodetection noise jointly constrain the attainable signal-to-noise-plus-distortion ratio and communication bandwidth. For the system studied here, assuming a flat and broad instantaneous frequency response, the model predicts a maximum quadrature phase-shift keying data rate of approximately $5~\mathrm{Mbps}$ for an error vector magnitude below $17.5\%$. Thus, nonlinear distortion can limit the usable communication bandwidth well below the large instantaneous bandwidths reported in the literature.
\end{abstract}

\maketitle


\section{Introduction}

\begin{figure}[!h]
    \centering
    \includegraphics[width=\linewidth]{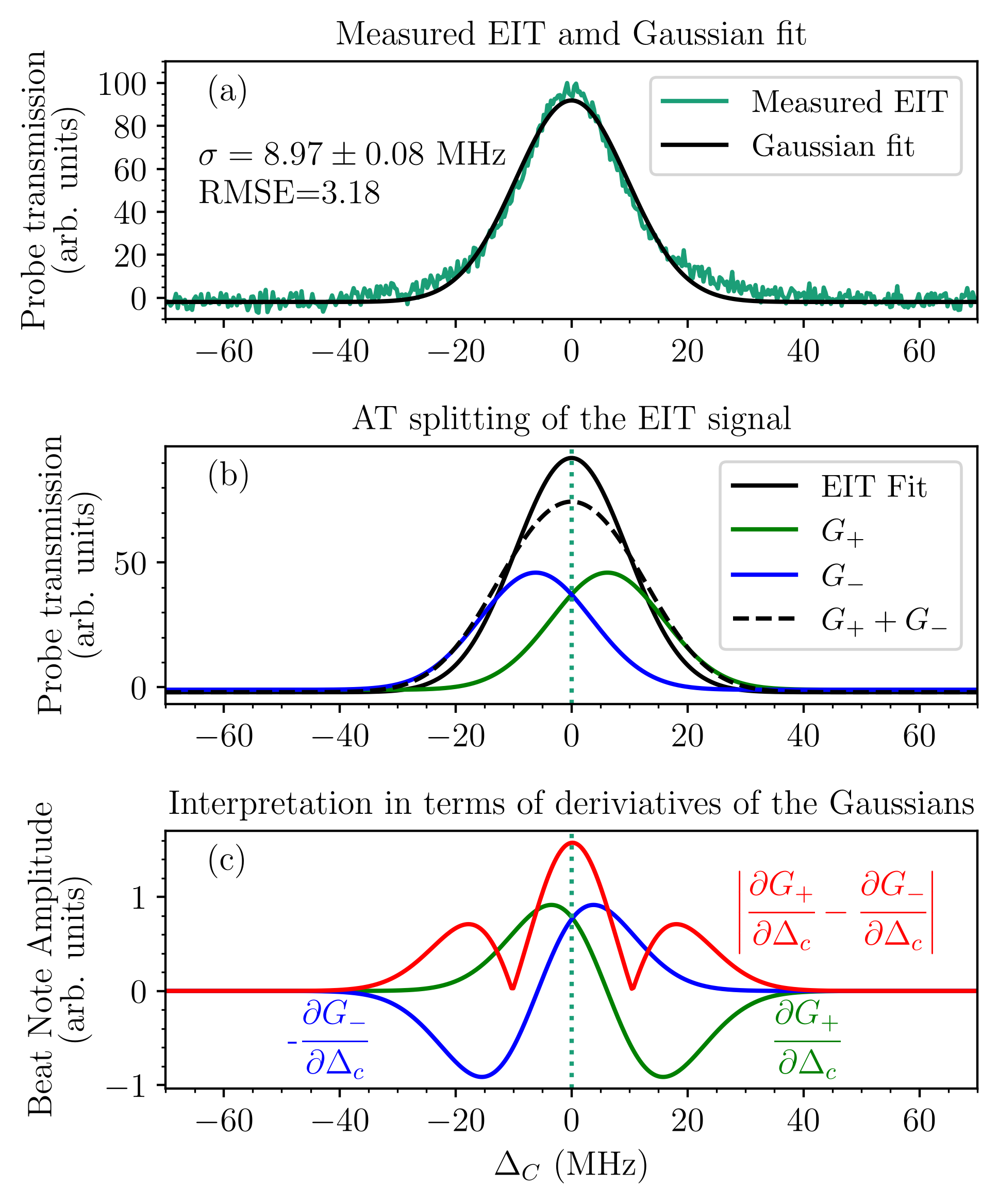}
    \caption{(a) Measured EIT spectrum with the corresponding Gaussian fit overlaid. The fitted Gaussian standard deviation is $\sigma = (8.97\pm0.08)~\mathrm{MHz}$. See Supplementary Information for the details of the fit. (b) Illustration of the AT splitting of the EIT resonance. The AT-split spectrum is represented by two Gaussian components, denoted $G_+$ and $G_-$. (c) Illustration of the beat-note response as the combination of the slopes of the two Gaussian components forming the AT split EIT spectrum.}
    \label{fig:EIT_Gaussian_Interpretation}
\end{figure}

Rydberg atom-based sensors provide an atomic-scale equivalent to conventional receiving antennas for the detection and measurement of electromagnetic fields \cite{holloway_broadband_2014, sedlacek_microwave_2012}. Owing to their large electric-dipole moments and strong response to radio-frequency (RF) and microwave fields, Rydberg atoms can directly transduce incident electromagnetic fields into measurable changes in optical signal \cite{jiao_spectroscopy_2016}. In heterodyne operation, the Rydberg receiver additionally functions as an atomic mixer \cite{jing_atomic_2020,simons_rydberg_2019}. In this configuration, a local RF field serves as a local oscillator (LO) and coherently mixes with the incident signal field, enabling the extraction of both the amplitude and phase of the signal from the resulting optical beat note.

Although the heterodyne Rydberg-atom receiver has been widely investigated for wideband \cite{liu_continuous-frequency_2022,meyer_assessment_2020} and weak-field detection \cite{gordon_weak_2019, holloway_electric_2017}, sensing and communication applications \cite{fancher_rydberg_2021}, its nonlinear mixing characteristics and the associated performance limitations have received comparatively limited attention. In particular, the  atomic response can generate harmonic and intermodulation products, introduce signal compression, and ultimately limit important receiver metrics such as linearity, dynamic range, and signal fidelity. In a recent study, Goncalves et al.\cite{goncalves_nonlinearities_2026} experimentally investigated the harmonic and intermodulation components generated by the nonlinear response of a Rydberg-atom receiver and used a full atomic density-matrix model to describe their origin. However, the present work provides a simple and comprehensive theoretical description that directly relates the atomic electromagnetically induced transparency (EIT) and Autler-Townes (AT) spectral response to the generation of the fundamental and intermodulation components. This model is further extended to describe the response of the receiver to a quadrature-phase-shift-keying (QPSK) waveform.

For broadband signals, the intermodulation products that fall within the signal band contribute directly to in-band distortion, reducing the effective signal-to-noise-plus-distortion ratio (SNDR). Nonlinear distortion can also lead to spectral regrowth, adjacent-channel leakage, and an increase in the effective distortion floor. Intermodulation distortion therefore is an important consideration for broadband operation. Unlike the conventional electronic mixers, whose nonlinearities arise from semiconductor device physics and non-ideal switching dynamics \cite{shabanzadeh_circuit_2026}, the nonlinear response of a Rydberg receiver originates from its underlying atomic spectroscopic response. This nonlinear behavior can depend strongly on the LO and signal amplitudes, optical linewidth, and laser detuning. Quantifying these nonlinear effects is therefore essential for determining the operating range over which a Rydberg receiver can preserve signal fidelity.

In this work, we investigate how the nonlinear response of a Rydberg-atom receiver translates into its intermodulation and communication performance. For this, the EIT lineshape is approximated as a Gaussian and its standard deviation $\sigma$, shown in Fig.~\ref{fig:EIT_Gaussian_Interpretation}(a), is used as the characteristic parameter describing the nonlinear response. Using two-tone measurements over a broad range of LO and signal amplitudes, we experimentally characterize the fundamental and higher-order intermodulation responses and compare them with a nonlinear model of the Rydberg-atom receiver, obtaining good agreement over the measured parameter space. We show that the useful signal response is confined to a finite region of LO and signal amplitudes, within which the relative strengths of the fundamental and intermodulation components vary substantially.

Like conventional receivers, the performance of Rydberg-atom receivers is often characterized by bandwidth.  Unfortunately, the term {\it bandwidth} is used in different contexts within the Rydberg-atom sensor community, which can make its interpretation ambiguous.  Thus, it is useful to distinguish among three different bandwidth terms.  The first is the {\it tuning bandwidth}, which defines the range of carrier frequencies over which a Rydberg receiver can be tuned to detect an RF signal. The tuning range can extend from the MHz regime \cite{liu_highly_2022,talashila_characteristics_2026, hammerland_tuning_2026} to hundreds of GHz \cite{thaicharoen_electromagnetically_2019} and, depending on the sensing approach, toward the THz regime \cite{lin_terahertz_2025}. The second term is the {\it instantaneous bandwidth}, which defines the maximum frequency span over which RF components can be detected simultaneously without retuning the receiver. Instantaneous bandwidths of tens of MHz range in a single-channel superheterodyne receiver have been  experimentally demonstrated \cite{yang_highly_2024}. More recently, channelized architectures based on spatially multiplexed Rydberg detection have demonstrated instantaneous bandwidths exceeding $200~\mathrm{MHz}$ \cite{jiao_-demand_2026}. The third term is the {\it communication bandwidth}, which describes the bandwidth over which information-bearing signals can be received with a specified communication-performance criterion, such as error-vector magnitude (EVM) or bit-error rate (BER). Equivalently, it determines the range of symbol or bit rates that can be supported for a specified modulation format. Let us look at the three types of bandwidths in more detail.

Rydberg-atom sensors can, in general, be tuned over a broad range of microwave frequencies by selecting appropriate Rydberg states and transitions for RF electrometry using Autler--Townes (AT) splitting. While the accessible carrier frequencies are determined by the atomic excitation scheme and the available laser wavelengths, heterodyne Rydberg sensing exhibits a finite intermediate-frequency (IF) response bandwidth \cite{jing_atomic_2020}, which determines the range of modulation or beat frequencies that can be detected without retuning the receiver. This IF response is closely related to the dynamical response of the atomic system, although the experimentally measured bandwidth can also be influenced by the optical and electronic detection chain. The achievable communication bandwidth, however, is distinct from this IF bandwidth. It is determined not only by the frequency-dependent receiver response, but also by integrated detection noise and nonlinear distortion generated within the receiver. Consequently, a Rydberg receiver may possess a relatively broad IF response while supporting a smaller communication bandwidth for a specified modulation format and EVM or BER requirement.

The nonlinear response of the receiver primarily places constraints on the achievable communication bandwidth. For a desired signal power ($P_S$), noise spectral density ($N_0$), in-band nonlinear distortion power ($P_D$), and a specified EVM requirement ($\mathrm{EVM_{req}}$), the allowable receiver noise bandwidth can be expressed as (see Supplementary Information for detailed derivation)

\begin{equation}
    B_n \le \left[ \mathrm{EVM_{req}^2 - \frac{P_D}{P_S}} \right] \frac{P_S}{N_0}.
\end{equation}

By combining the measured fundamental signal, in-band intermodulation components, and photodetection noise floor, we determine the resulting SNDR and compare it with the EVM response of a QPSK signal.  The SNDR and EVM are evaluated across the parameter space of LO and signal powers. The nonlinear model is then used to reproduce the fundamental, intermodulation products, SNDR and  EVM responses, showing good agreement with the experimental results. Although the two-tone measurements are not fully sufficient to represent the response of the receiver to a modulated waveform such as QPSK, they clearly demonstrate the degradation caused by nonlinear distortion. Finally, the model is used to predict the achievable communication bandwidth for an atomic receiver with a larger IF bandwidth, where the limitation is determined mainly by the combined effects of noise and nonlinear distortion, rather than by the intrinsic frequency response of the receiver.
 

 Several studies have investigated the channel capacity of Rydberg-atom receivers. Cox  et al. introduced a quantum limit for the data capacity of an atomic receiver and experimentally demonstrated data reception over bandwidths extending from kHz to tens of MHz \cite{cox_quantum-limited_2018}. Otto  et al. investigated the scaling of Shannon channel capacity using a spatially distributed array of Rydberg-atom receivers \cite{otto_data_2021}. Nowosielski  et al. experimentally demonstrated, with a three-photon optical excitation, the reception of quadrature amplitude modulated signals using a warm Rydberg-atom heterodyne receiver, analyzed communication error for different modulation formats, and reported an achievable single-channel capacity of $19.3~\mathrm{Mbps}$ \cite{nowosielski_warm_2024}. More recently, Qu and Zhang developed a dynamical model of a superheterodyne Rydberg receiver and optimized its channel capacity considering the receiver  frequency response and noise contributions \cite{qu_channel-capacity_2026}, while Zhu and Dai developed a dynamic signal model and investigated the corresponding capacity limit from the atomic transfer function and noise processes \cite{zhu_general_2026}.  Manchaiah et al. recently examined the bandwidth of a Rydberg-atom receiver using modulated signals and showed, through EVM measurements at increasing symbol rates, that the communication bandwidth can differ substantially from the pure-tone response bandwidth because the modulated signal occupies a finite spectral range and accumulates noise across that bandwidth \cite{manchaiah_probing_2026}. These previous capacity studies were primarily based on the receiver bandwidth, signal-to-noise ratio (SNR), or linearized atomic response. Although nonlinear effects have been considered in some cases to identify the range of validity of the linear response, the influence of experimentally measured intermodulation distortion on communication performance has not been explicitly included in the channel-capacity analysis. In the present work, we directly characterize the nonlinear distortion of the Rydberg receiver and investigate its consequence for communication performance through the measured EVM over a wide range of parameters of signal and LO powers. While a complete channel-capacity analysis is beyond the scope of the present work, the observed degradation of EVM due to nonlinear distortion has direct implications for the achievable information rate and consequently, the practical channel capacity of the Rydberg receiver.

The rest of the paper is organized as  follows. Section~\ref{sec:Exp_Setup} describes the experimental setup and the relevant parameters. Section~\ref{sec:Gaussian_Interpretation} develops the mathematical description of the nonlinear response of the Rydberg atom receiver and compares the measured results with the corresponding model predictions. Section~\ref{sec:Nonlinear_EVM} evaluates the effect of these nonlinear components on the EVM response.  Building on these results, section~\ref{sec:Bandwidth} presents  the attainable communication bandwidth and its limitations arising from nonlinear distortion and noise. The broader implications and limitations of this work are discussed in section~\ref{sec:Discussion}. The Supplementary Information provides additional details on the experimental-system characterization, mathematical derivations, EIT lineshape fitting, EVM results, and the derivation of the intermodulation null conditions.

\begin{figure}[!h]
    \includegraphics[width=0.99\linewidth]{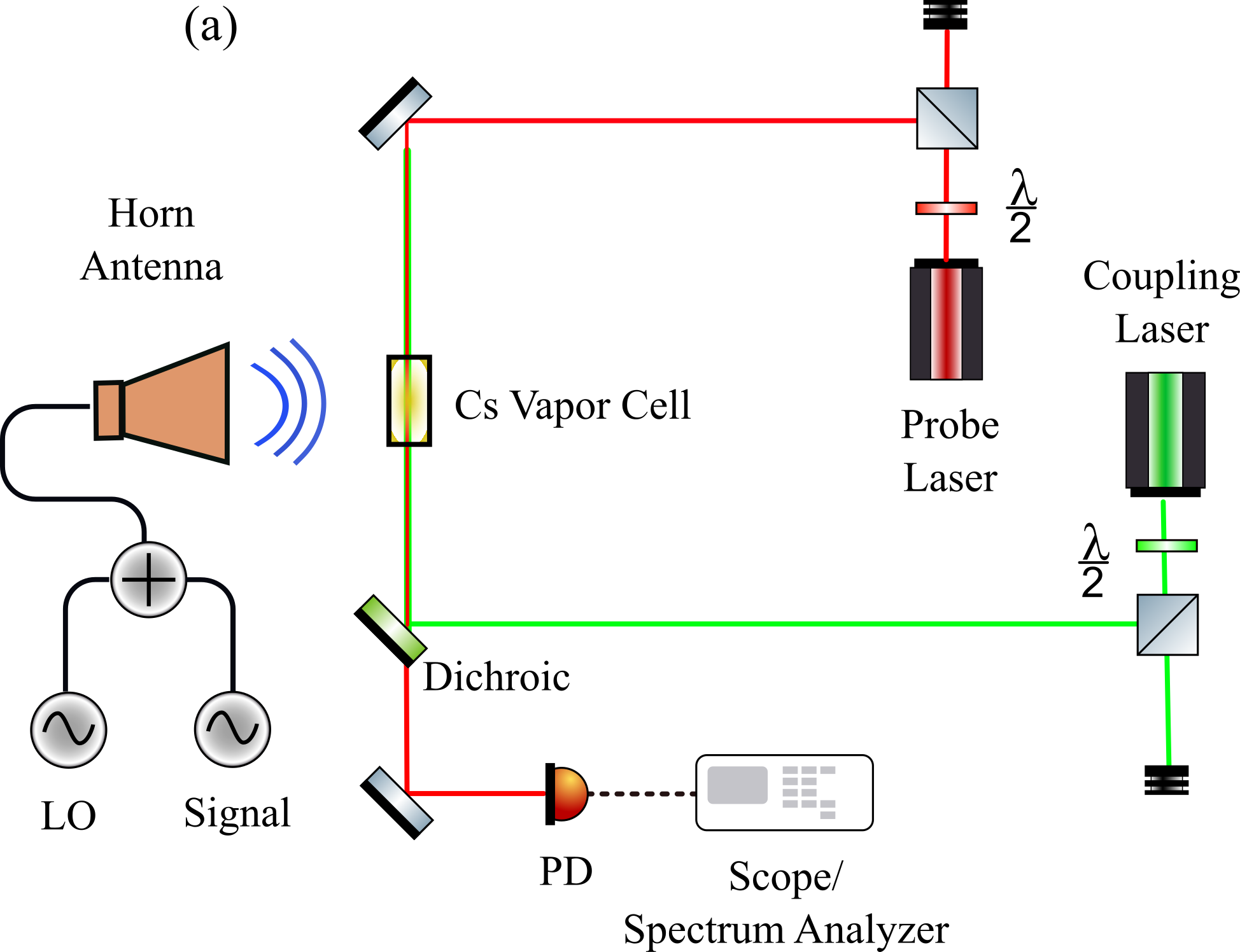}
    \includegraphics[width=0.7\linewidth]{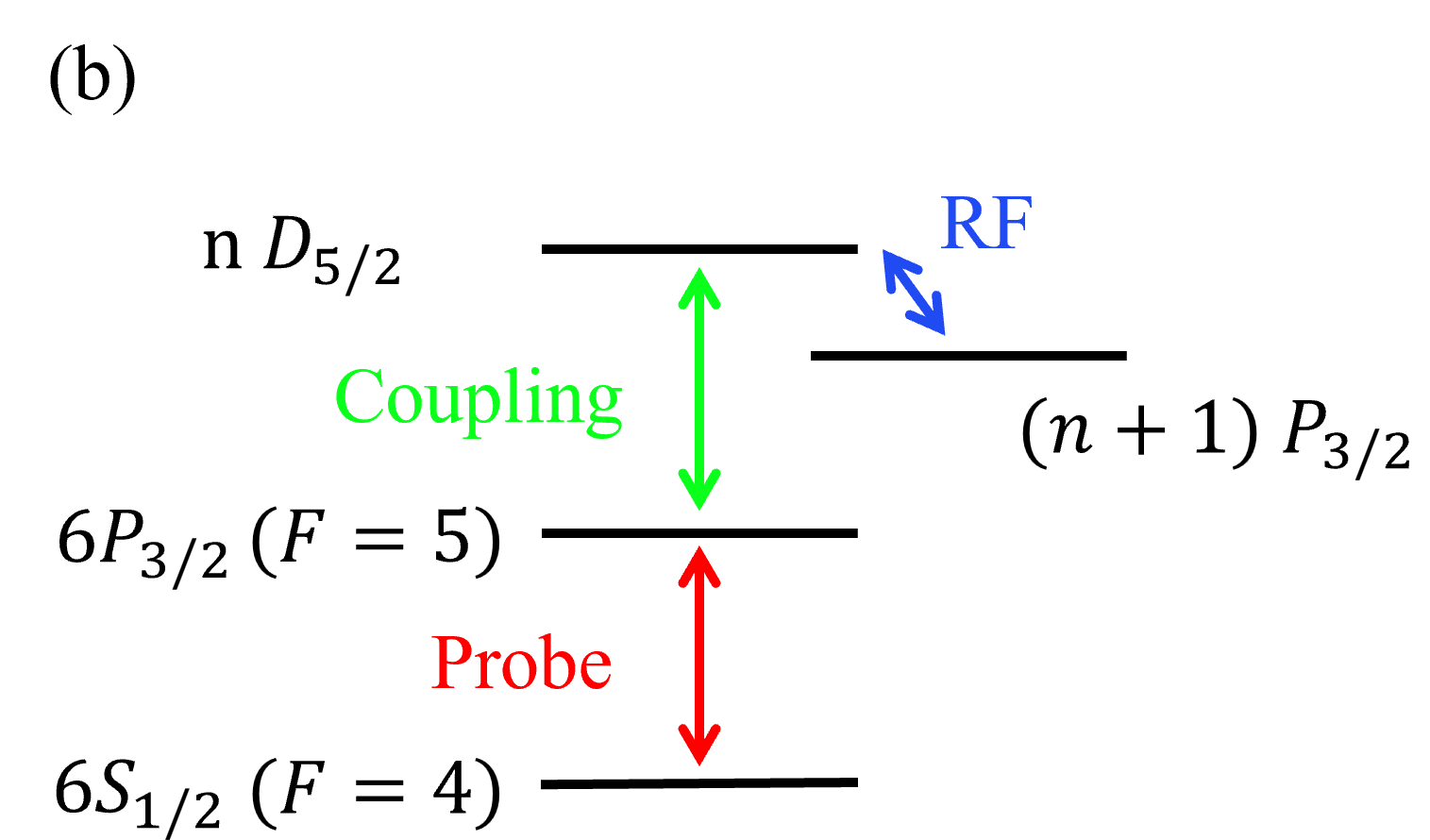}
    \caption{(a) Experimental setup used in this study. The cesium-vapor-filled cylindrical cell is excited by counter-propagating coupling and probe beams. Two RF sources, the LO and signal, used for heterodyne detection are combined and applied on the vapor cell through a horn antenna for resonant RF excitation. The two lasers are linearly polarized and co-polarized with the RF field. All the electrical-field polarization vectors are perpendicular to the plane of the  paper. (b) Two-photon Rydberg excitation scheme with a resonant RF field coupling two Rydberg states. $n$ denotes the principal quantum number. In this work, $n=50$, corresponding to a resonant RF transition frequency of $5709\,\mathrm{MHz}$.}
    \label{fig:Schematic_Levels}
\end{figure}
\section{Experimental setup}
\label{sec:Exp_Setup}

In the two-photon Rydberg-excitation scheme shown in Fig.~\ref{fig:Schematic_Levels}(b), counter-propagating probe and coupling laser beams prepare a target Rydberg state in cesium atoms contained within a room-temperature, cylindrical vapor cell with $25~\mathrm{mm}$ diameter  and $45~\mathrm{mm}$ length. The probe beam addresses the D2 transition ($6S_{1/2}\leftrightarrow6P_{3/2}$), while the coupling beam drives the $6P_{3/2}\leftrightarrow50D_{5/2}$ transition to prepare the atoms in the $50D_{5/2}$ Rydberg state. The coupling and probe beams have approximate full width at half maximum (FWHM) diameters of $130~\mathrm{\mu m}$ and $140~\mathrm{\mu m}$, respectively, with corresponding optical powers of $130~\mathrm{mW}$ and $56~\mathrm{\mu W}$. Both laser beams are (linearly) co-polarized with the RF fields.

The measured EIT spectrum in the absence of an applied RF field is presented in Fig.~\ref{fig:EIT_Gaussian_Interpretation}(a). The corresponding Gaussian fit gives a standard deviation of $\sigma=(8.97\pm0.08)~\mathrm{MHz}$. A representative AT-split EIT spectrum is shown in Fig.~\ref{fig:EIT_Gaussian_Interpretation}(b). The individual Gaussian components associated with the AT splitting are denoted by $G_{\pm}$ in the plot. The signal and LO electric-field amplitudes experienced by the atoms are independently calibrated using the measured AT splitting. Details of the experimental-system characterization, including the selection of laser detunings, RF-field calibration, IF bandwidth, and sensitivity measurements, are provided in the Supplementary Information.

The resonant RF field addresses the $50D_{5/2}\leftrightarrow51P_{3/2}$ transition at a frequency of $5709~\mathrm{MHz}$. The electric-dipole matrix element for the $\ket{50D_{5/2},m_J=1/2}\leftrightarrow\ket{51P_{3/2},m_J=1/2}$ transition, corresponding to $\pi$-polarized RF excitation ($\Delta m_J=0$), is calculated using the Alkali Rydberg Calculator (ARC) \cite{sibalic_arc_2017}  to be $d=1644.57~ea_0$. Throughout this work, the RF Rabi frequency $\Omega$ is expressed in frequency units (Hz) rather than angular frequency units (rad/s). The Rabi frequencies associated with the signal and LO RF fields are related to their respective electric-field amplitudes through  $\Omega = \vec{d} \cdot \vec{E}_{\mathrm{RF}}/ h$, where $\vec{d}$ is the transition dipole matrix element, $\vec{E}_{\mathrm{RF}}$ is the peak RF electric field amplitude vector, and $h$ is Planck's constant.
\begin{figure*}
    \centering
    \includegraphics[width=0.99\linewidth]{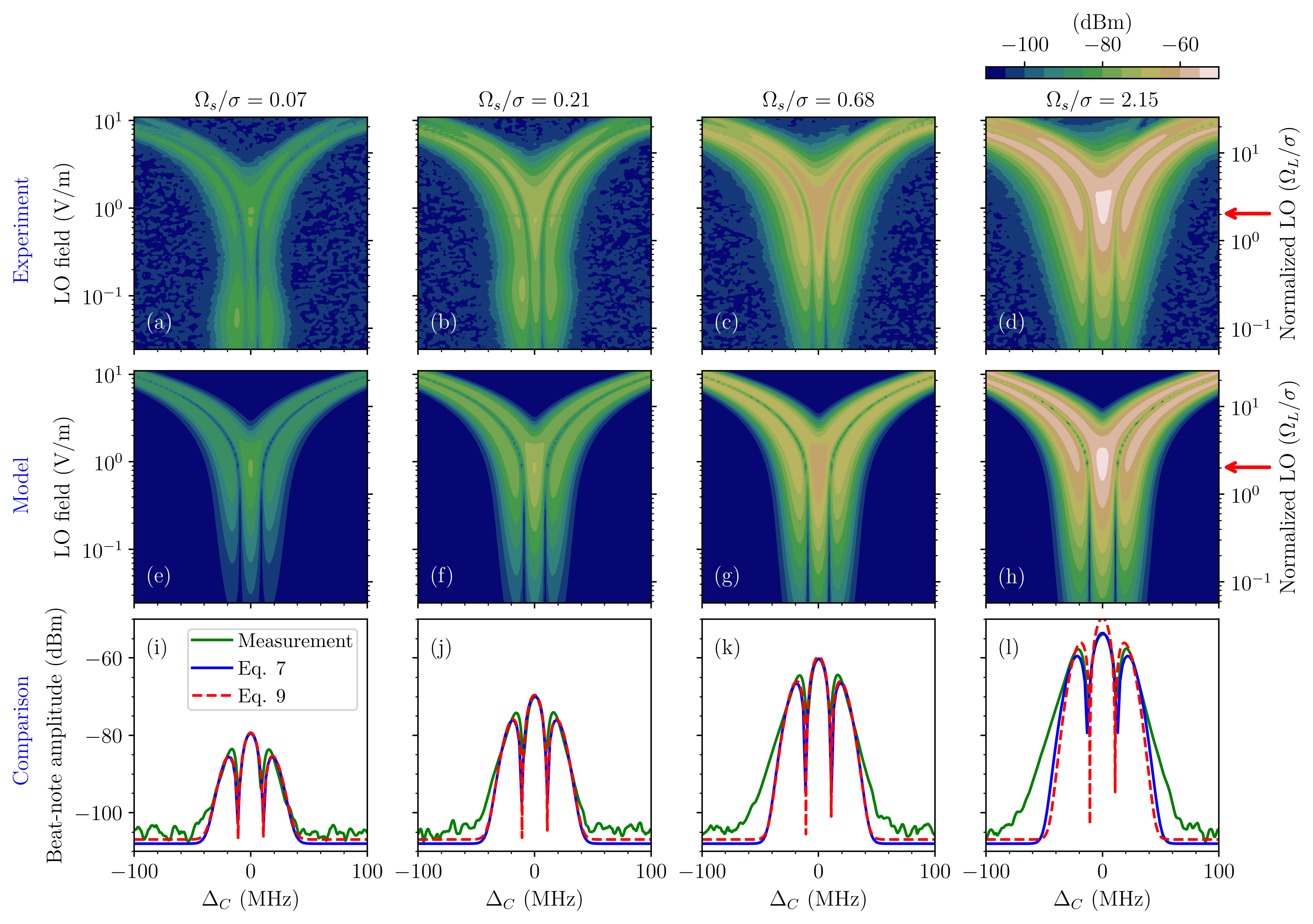}
    \caption{(a-d) Experimentally obtained beat-note amplitude as a function of coupling-laser detuning and LO field amplitude for normalized signal amplitudes of $\Omega_S/\sigma = 0.07$, $0.21$, $0.68$, and $2.15$, respectively. (e-h) Corresponding beat-note amplitudes calculated using the model described by Eq.~\ref{eq:Beat_note_in_terms_of_Rabi_rate}. (i-l) Comparison of the experimental and modeled line cuts of the beat-note amplitude at a fixed LO electric-field amplitude of $0.86~\mathrm{V/m}$ ($\Omega_L/\sigma\approx2$) for the corresponding signal amplitudes shown in (a-d). The LO amplitude used for the line-cut comparison is indicated by red arrows in panels (d) and (h).}
    \label{fig:Beatnote_differenatiation_2D}
\end{figure*}
\section{Gaussian interpretation of the nonlinearity}
\label{sec:Gaussian_Interpretation}
\subsection{Time-dependent EIT response under heterodyne excitation}

When the Rydberg-atom receiver is subjected to heterodyne excitation, the EIT signal is modulated in time, and the beat note component at difference frequency carries the information of interest. First, without making any assumptions regarding the relative amplitudes of signal and LO fields, this subsection describes the full time-dependent EIT response through the instantaneous Autler-Townes (AT) splitting. A simplified description is then presented in the weak-signal limit, in which the beat-note response is expressed in terms of the derivatives of the two Gaussian components forming the AT-split EIT spectrum. Based on this model, the predicted beat note response is compared with the corresponding experimental measurements. 

When broadened by inhomogeneous mechanisms, the measured EIT spectrum is well-approximated by a Gaussian profile \cite{holloway_autlertownes_2026}, which we express as
\begin{equation}
    EIT(\Delta_C) = \exp\left[-\frac{1}{2\sigma^2}  \Delta_C^2\right],
\end{equation}
where $\Delta_C$ is the detuning of the coupling laser frequency and $\sigma$ is the standard deviation of the Gaussian fit of the EIT resonance. Applying a coherent RF source of the form $E_{0} \hat{\epsilon} \exp\left[-i(\omega t + \phi)\right]$ with corresponding Rabi rate $\Omega$, the resulting Autler-Townes (AT) split EIT spectrum can be expressed as a sum of two Gaussians separated by the Rabi frequency $\Omega$. Mathematically, it can be expressed as
\begin{equation}
    EIT_{AT} (\Delta_C,\Omega) = \cfrac{1}{2} \left[G_{+}(\Delta_C,\Omega) + G_{-}(\Delta_C,\Omega)\right],\label{eq:EIT_in_Gaussians_PM}
\end{equation}
where
\begin{equation}
     G_{\pm}(\Delta_C,\Omega) = \exp\left[ -\frac{1}{2\sigma^2}\left(\Delta_C \mp \frac{\Omega}{2}\right)^2\right], \label{eq:Gausians_PM}
\end{equation}
and the separation between the peaks of the two Gaussians $G_{\pm}(\Delta_C,\Omega)$  is given by $\Omega$. 

\begin{figure}
    \centering
    \includegraphics[width=\linewidth]{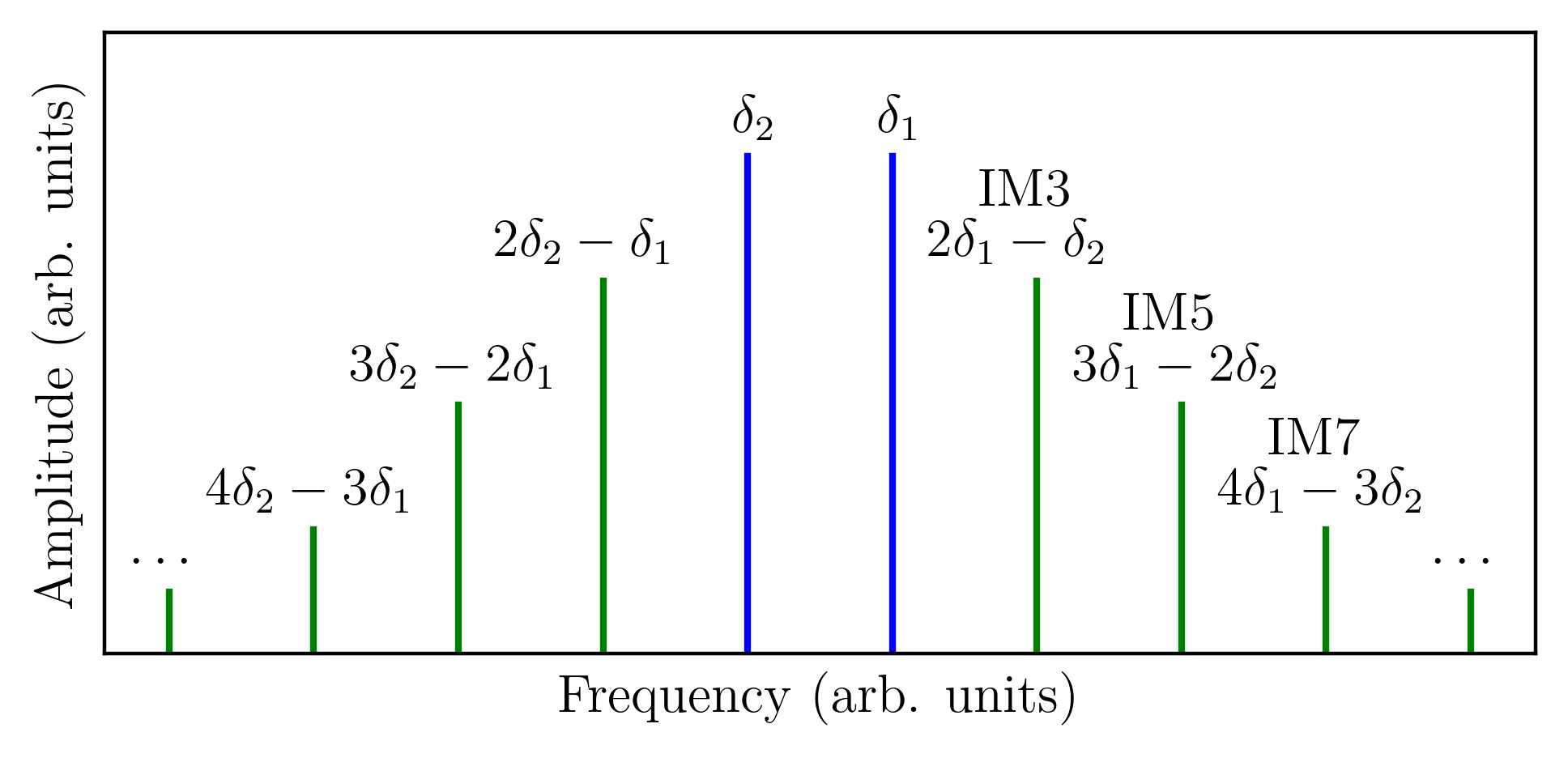}
    \caption{Representative illustration of intermodulation components generated by a nonlinear system in response to a two-tone input. }
    \label{fig:IM_Representative}
\end{figure}

In heterodyne reception, an RF field serves as a LO reference enabling  extraction of the beat note frequency and the relative phase of the signal RF field. Hence, the instantaneous Rabi rate $\Omega$ contains  contributions from both the signal and LO RF fields. Working in the rotating frame defined by the LO field, the complex Rabi frequency phasor can be written as 
\begin{equation}
    \tilde\Omega(t) = \Omega_L +\Omega_{S}e^{i(\delta t + \phi)},
\end{equation}
and so the effective instantaneous magnitude is
\begin{equation}
    \Omega(t) = |\tilde\Omega(t)| = \sqrt{\Omega_L^2 +\Omega_{S}^2 + 2 \Omega_L \Omega_S \cos (\delta t + \phi)},
    \label{eq:Instaneous_Rabi}
\end{equation}
where $\Omega_L$ and $\Omega_S$ are the Rabi frequencies associated with the LO and the signal fields, respectively. The  angular beat frequency is defined as $\delta = \omega_L-\omega_S$ where, $\omega_L$ and $\omega_S$ correspond to the angular frequencies of the LO and the signal fields, respectively. Hence, the resulting time-dependent AT response described by eq.~\ref{eq:EIT_in_Gaussians_PM} can be expressed as

\begin{align}
    E&IT_{AT}(\Delta_C,\Omega_S,\Omega_L,t) \nonumber\\
    &= \exp\left[ -\frac{\Delta_C^2}{2\sigma^2}\right] \exp\left[ -\frac{\Omega^2(t)}{8\sigma^2}\right]
    \cosh\left[ \frac{\Delta_C\Omega(t)}{2\sigma^2}\right] \nonumber  \\
    &=  \exp\left[ -\frac{\Delta_C^2}{2\sigma^2}\right] 
    \exp\left[ -\frac{\Omega_L^2+\Omega_S^2+2\Omega_S\Omega_L\cos(\delta t+\phi)}{8\sigma^2}\right] \nonumber \\
    &\quad\times\cosh\left[ \frac{\Delta_C}{2\sigma^2} \sqrt{\Omega_L^2 +\Omega_{S}^2 + 2 \Omega_L \Omega_S \cos (\delta t + \phi)}\right] .\label{eq:Beat_note_in_terms_of_Rabi_rate}
\end{align}
While this expression formally represents the time modulated EIT lineshape in the presence of LO and signal fields, it is more insightful here to interpret the result as a beat-note amplitude atop an AT-split profile induced by the LO. For the weak signal approximation $\Omega_S \ll \Omega_L$, the effective instantaneous magnitude of the Rabi frequency can be expressed as \cite{simons_rydberg_2019,jing_atomic_2020}
\begin{equation}
    \Omega(t) \simeq \Omega_L + \Omega_S \cos(\delta t + \phi).
\end{equation}
With this approximation, the beat-note amplitude, corresponding to the difference-frequency component of the time-dependent EIT response, can be expressed as  ( see Supplementary Information for the derivation)
\begin{align}
B(\Delta_C,\Omega_L,\Omega_S,t) &\simeq -\frac{\Omega_S}{4} \cos(\delta t + \phi)
\left[ \frac{\partial G_+ }{\partial\Delta_C}  - \frac{\partial G_-}{\partial\Delta_C}  \right],\\ \label{eq:EIT_Differentiation}
&\simeq \frac{\Omega_S}{2}\cos(\delta t + \phi)
\left[ \frac{\partial G_+ }{\partial\Omega_L}  + \frac{\partial G_-}{\partial\Omega_L}  \right].
\end{align}

\begin{figure*}
    \centering
    \includegraphics[width=\linewidth]{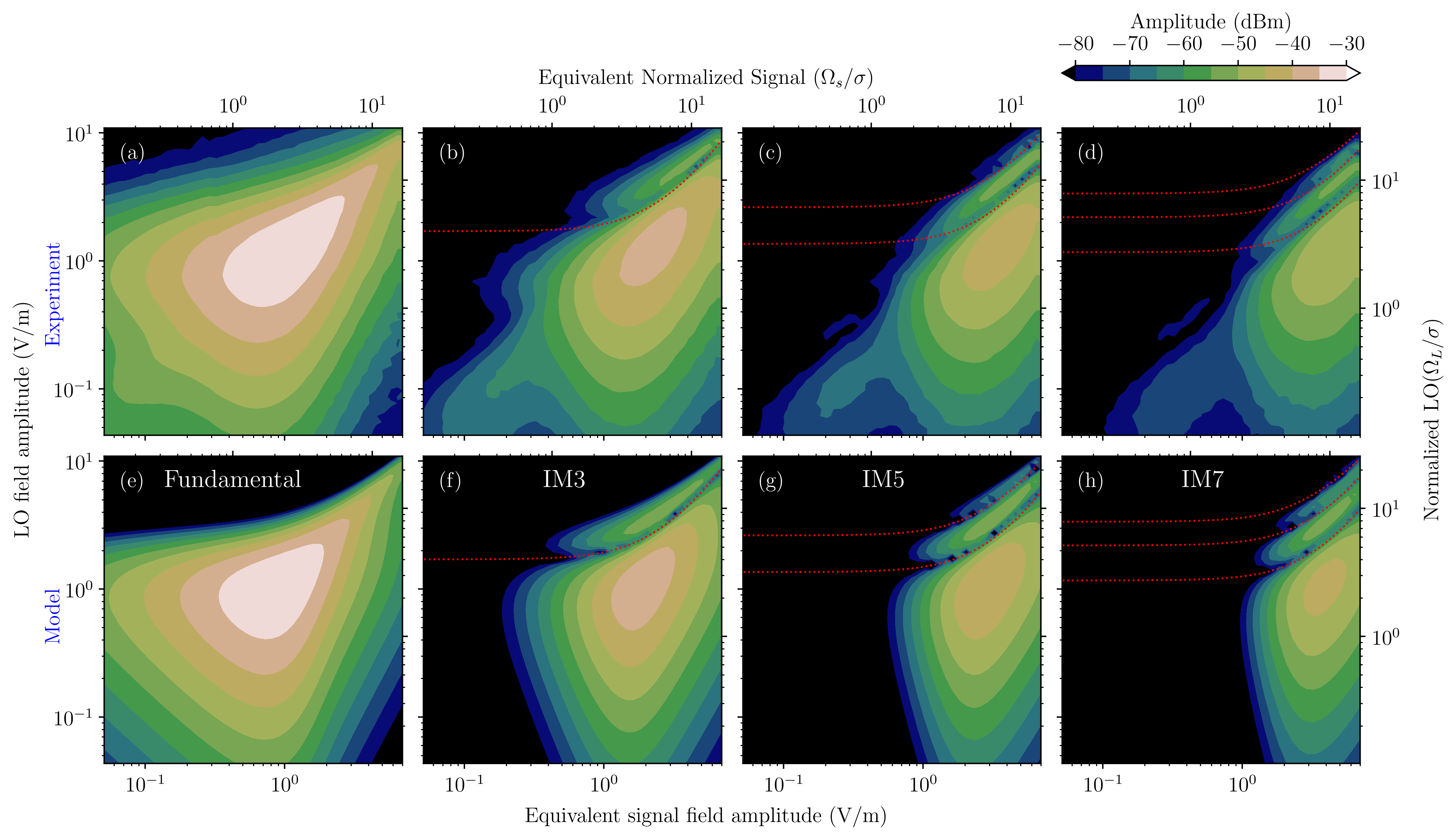}
    \caption{ (a-d) Measured amplitudes of the fundamental, third-order intermodulation (IM3), fifth-order intermodulation (IM5), and seventh-order intermodulation (IM7) components, respectively. (e-h) Corresponding modeled amplitudes of the fundamental, IM3, IM5, and IM7 components, respectively. Each quantity is plotted in the two-dimensional parameter space defined by the signal and LO field amplitudes. The absolute signal electric-field amplitude at the atoms, expressed in $\mathrm{V/m}$, is shown on the lower $x$-axis, while the corresponding signal normalized Rabi amplitude $\Omega_S/\sigma$ is shown on the upper $x$-axis. Similarly, the absolute LO electric-field amplitude is shown on the left $y$-axis, while the corresponding normalized Rabi amplitude $\Omega_L/\sigma$ is shown on the right $y$-axis. The red dotted lines overlaid on the corresponding panels indicate the analytically determined null conditions for the IM3, IM5, and IM7 components. This representation enables direct comparison between the experimentally calibrated electric-field amplitudes and the dimensionless parameters used in the nonlinear model.  }
    \label{fig:Fundamental_Harmonics_2D}
\end{figure*}

\begin{figure*}
    \centering
    \includegraphics[width=\linewidth]{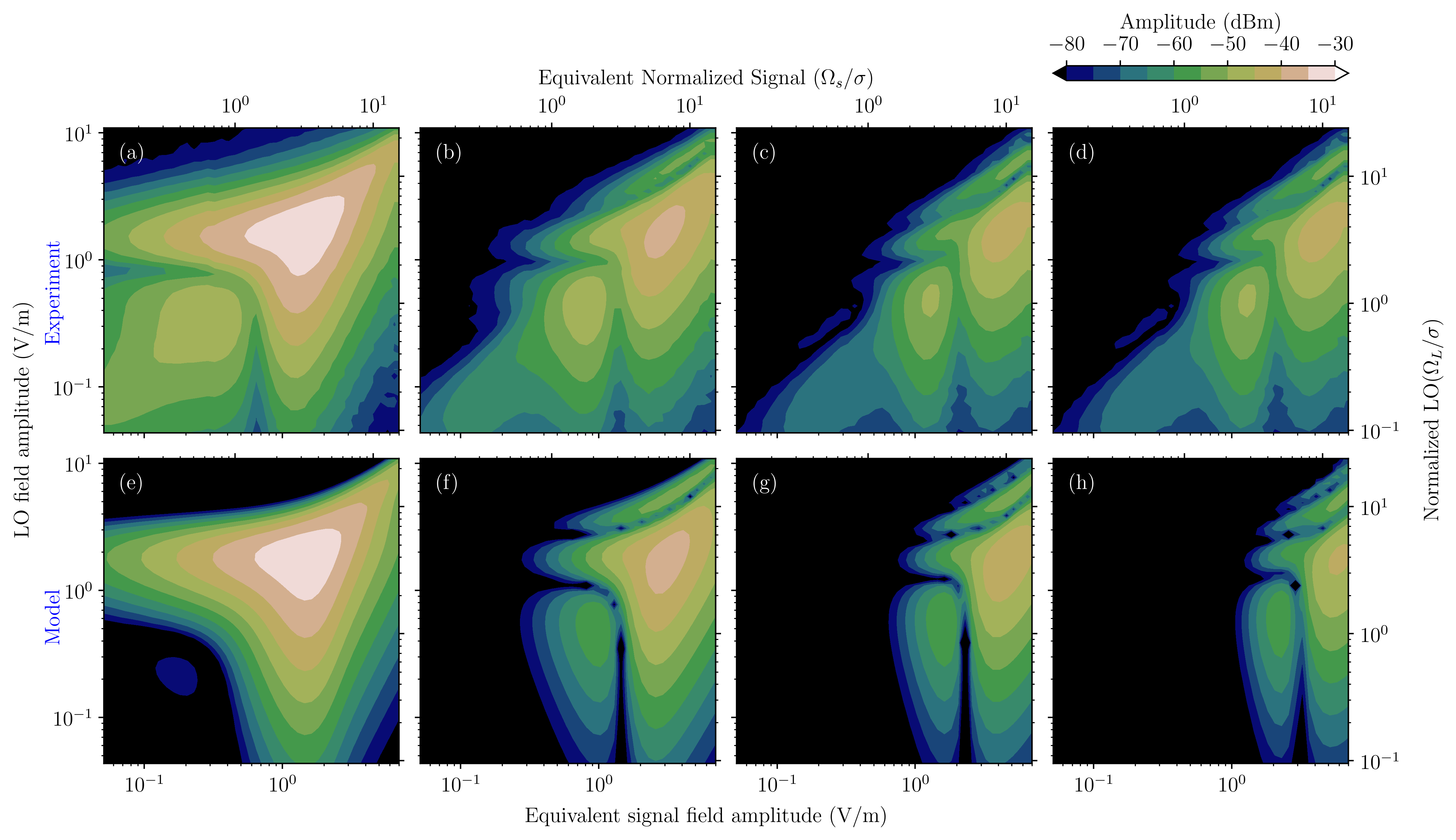}
    \caption{The same quantities and panel organization as in Fig.~\ref{fig:Fundamental_Harmonics_2D} are shown for a coupling-laser detuning of $-10~\mathrm{MHz}$. }
    \label{fig:Beatnote_differenatiation_2D_detuned}
\end{figure*}

From this relation, the beat-note amplitude as a function of coupling-laser detuning is determined by the difference between the slopes of the two Gaussian components forming the AT-split EIT spectrum. A representative illustration showing this relation is presented in Fig.~\ref{fig:EIT_Gaussian_Interpretation}(c). Further, the experimentally obtained beat-note amplitude as a function of coupling-laser detuning and LO amplitude, for various signal amplitudes, is presented in Fig.~\ref{fig:Beatnote_differenatiation_2D}(a-d), while the corresponding modeled response calculated using Eq.~\ref{eq:Beat_note_in_terms_of_Rabi_rate} is shown in Fig.~\ref{fig:Beatnote_differenatiation_2D}(e-h). The corresponding experimental and modeled line cuts of both Eq.~\ref{eq:Beat_note_in_terms_of_Rabi_rate} and Eq.~\ref{eq:EIT_Differentiation} at a fixed LO amplitude of $0.86~\mathrm{V/m}$ ($\Omega_L/\sigma\approx 2$) are compared in Fig.~\ref{fig:Beatnote_differenatiation_2D}(i-l). At low signal and LO amplitudes, deviations between the model and the experiment are observed, where the AT splitting exhibits nonlinear behavior \cite{holloway_electric_2017, hao_rydberg_2019,hao_nonlinearity_2019}. Apart from these deviations in the weak-field regime, the agreement between results demonstrates that the EIT lineshape determines the heterodyne beat-note conversion over a substantial portion of the parameter space formed by $\Omega_s$, $\Omega_L$, and $\Delta_C$.

\subsection{Intermodulation distortion}

\begin{figure*}
    \centering
    \includegraphics[width=\linewidth]{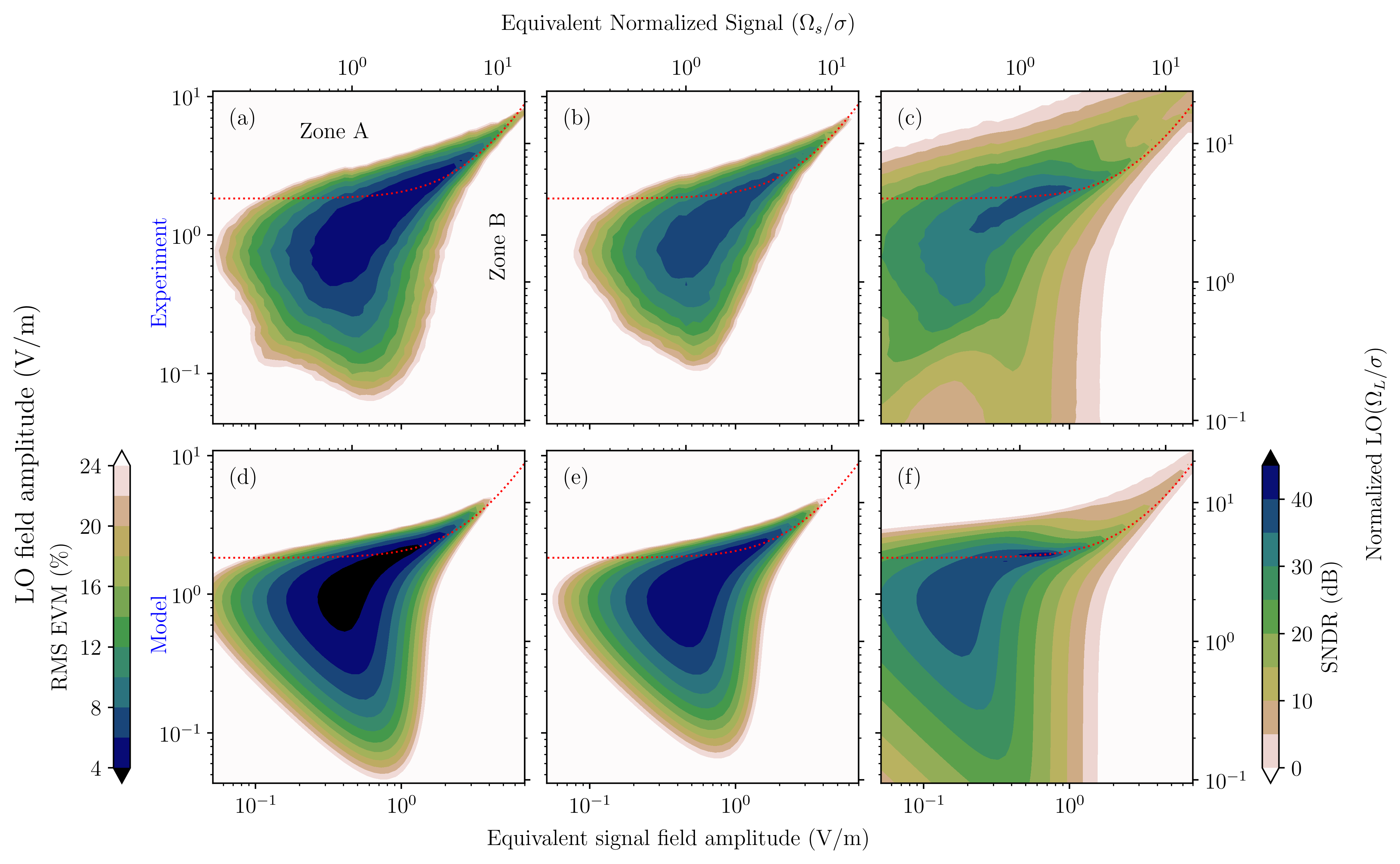}
    \caption{(a,b) Measured EVM as a function of LO and equivalent signal amplitude for a QPSK waveform at bit rates of $50~\mathrm{kbps}$ and $100~\mathrm{kbps}$, respectively, with a  beat-note frequency of $200~\mathrm{kHz}$. (c) SNDR calculated using the measured fundamental and IM3, IM5, and IM7 amplitudes shown in Fig.~\ref{fig:Fundamental_Harmonics_2D}(a-d). (d-f) Corresponding modeled  results for the quantities shown in panels (a-c), respectively. The IM3 null trajectory predicted by Eq.~\ref{eq:Expression_IM3_Null} is overlaid on all panels. The colorbar for the EVM plots in panels (a,b,d,e) is shown on the left, while that for the SNDR plots in panels (c,f) is shown on the right. }
    \label{fig:EVM_Experiment}
\end{figure*}

In active or passive nonlinear RF systems, multitone  excitation generates intermodulation products at linear combinations of the input frequencies. Odd-order intermodulation products are particularly important as they can fall within the band of interest and degrade the signal quality \cite{pedro_intermodulation_2003, lui_passive_1990}.  Even-order nonlinearities generally produce sum- and difference-frequency components, including components near DC and around even harmonics, which typically fall outside the frequency band of interest considered here. An illustrative example of odd-order intermodulation products generated by a nonlinear system under  two-tone excitation is shown in Fig.~\ref{fig:IM_Representative}. When these intermodulation products interfere with the desired signal, the resulting signal degradation is referred to as intermodulation distortion (IMD).

 To investigate the  intermodulation  response arising  from the nonlinear behavior of the Rydberg-atom receiver, we consider a signal composed of two frequency components. Taking the LO as the rotating frame reference, the combined effective Rabi frequency can be written as 
\begin{equation}
    \tilde\Omega(t) = \Omega_L +\Omega_{1}e^{i\delta_1 t} +\Omega_{2}e^{i\delta_2 t},
\end{equation}
where $\Omega_L$, $\Omega_1$ and $\Omega_2$ are the Rabi frequencies associated with the LO and the two signal components, respectively. The corresponding angular beat frequencies are 
\begin{equation}
    \delta_1 = \omega_L - \omega_1,  \qquad \delta_2 = \omega_L - \omega_2.
\end{equation}

The instantaneous AT splitting is determined by $\Omega(t)=|\tilde\Omega(t)|$, giving
\begin{align}
\Omega^2(t) &= \Omega_L^2 +\Omega_1^2 +\Omega_2^2 + 2 \Omega_L\Omega_1\cos\delta_1t + 2 \Omega_L\Omega_2\cos\delta_2t \nonumber \\
    &\quad+ 2 \Omega_1\Omega_2\cos[(\delta_2-\delta_1)t]. \label{eq:Instantaneous_Rabi_Two_Tone}
\end{align}

This instantaneous Rabi rate contains contributions from the LO and the two signal components at their corresponding beat note frequencies. Substituting this expression into eq.\ref{eq:Beat_note_in_terms_of_Rabi_rate}, and considering the case of zero coupling detuning $\Delta_C=0$, yields the resulting time dependent EIT response

\begin{align}
 \mathrm{EIT}_0(t) &= \exp\left[ -\frac{\Omega_L^2+\Omega_1^2+\Omega_2^2}{8\sigma^2}\right] \nonumber\\
    &\quad\times\exp\left[-\frac{\Omega_1\Omega_L}{4\sigma^2}\cos\delta_1 t\right]\exp\left[-\frac{\Omega_2\Omega_L}{4\sigma^2}\cos\delta_2 t\right]\nonumber\\&\quad\times\exp\left[-\frac{\Omega_1\Omega_2}{4\sigma^2}\cos((\delta_2-\delta_1) t)\right].\label{eq:Beat_Note_Nonlinear_zero_detuning}
\end{align}

 This nonlinear EIT response  generates mixing products at linear combinations of the two beat frequencies, $\delta_1$ and $\delta_2$. In particular, the odd-order intermodulation products appearing near the fundamental components are separated from them by integer multiples of the spacing $\delta_2-\delta_1=\omega_1-\omega_2$, as illustrated schematically in Fig.~\ref{fig:IM_Representative}. 
 
To experimentally characterize this nonlinear response, the Rydberg-atom receiver is subjected to a two-tone test. The signal consists of two equal amplitude frequency components separated by $20~\mathrm{kHz}$ and centered around a beat note frequency of  $200~\mathrm{kHz}$. By varying the applied signal and LO RF powers, the receiver response is measured using a spectrum analyzer and  the resulting intermodulation products are recorded.
 
 The experimentally obtained fundamental ($210~\mathrm{kHz}$), third-order intermodulation (IM3 : $230~\mathrm{kHz}$ ), fifth-order intermodulation (IM5 : $250~\mathrm{kHz}$), and seventh-order intermodulation (IM7 : $270~\mathrm{kHz}$) components  are plotted in Fig.~\ref{fig:Fundamental_Harmonics_2D}(a-d). The corresponding modeled responses, calculated using  Eq.~\ref{eq:Beat_Note_Nonlinear_zero_detuning} are shown in Fig.~\ref{fig:Fundamental_Harmonics_2D}(e-h). All panels share the same axes and are presented in terms of both the calibrated RF electric-field amplitudes and the corresponding normalized RF Rabi frequencies. To compare the modeled and measured amplitudes on the same scale, a single common offset is applied to the modeled data such that the peak modeled fundamental amplitude matches the corresponding experimental value; the same offset is then applied to all modeled intermodulation components. The measured noise floor is approximately $-80~\mathrm{dBm}$. The displayed color scale is extended below this level to preserve the full modeled response, although experimental values below the noise floor are not directly resolvable.

At low signal and LO amplitudes, deviations between the model and experiment are observed, while the overall trends are reproduced over the remaining parameter space. The purpose of this comparison is not to establish point-by-point quantitative agreement, but rather to identify the regions of signal and LO RF amplitudes over which the different intermodulation components emerge and become significant within the usable operating space of the receiver. 
 
 The intermodulation components exhibit distinct null trajectories that can be expressed as a relation between the normalized signal and LO amplitudes. The derivations of the approximate analytical conditions for cancellation of the IM3, IM5, and IM7 components are presented in the Supplementary Information, and the resulting condition for IM3  is
 
 \begin{equation}
    \left(\frac{\Omega_L}{\sigma}\right)_{\mathrm{IM3~Null}} \simeq \sqrt{16 + \frac{5}{3}\left(\frac{\Omega_S}{\sigma}\right)^2 + \frac{1}{18}\left(\frac{\Omega_S}{\sigma}\right)^4 }.
    \label{eq:Expression_IM3_Null}
\end{equation}

This predicted IM3-null trajectory is overlaid as a red dotted line on both the measured and modeled IM3 maps in Fig.~\ref{fig:Fundamental_Harmonics_2D}(b) and (f), respectively,  showing good agreement with the locations of the observed minima. Similar calculated null trajectories for the IM5 and IM7 components are overlaid  on their corresponding measured and modeled panels in Fig.~\ref{fig:Fundamental_Harmonics_2D}. The agreement in the locations and overall behavior of the null trajectories across these different intermodulation orders supports the ability of this simple model to capture the nonlinear response of the Rydberg-atom receiver.

From the above analysis, the fundamental beat note component exhibits signs of saturation with increasing signal and LO amplitudes. In the weak signal limit, an analytical condition for the  LO amplitude that maximizes the fundamental beat-note amplitude can be obtained, providing an optimum operating point for the Rydberg-atom receiver. At zero  coupling detuning, $\Delta_C=0$, and for $\Omega_S\ll\Omega_L$, the fundamental beat-note amplitude obtained from Eq.~\ref{eq:Beat_note_in_terms_of_Rabi_rate} is 
\begin{align}
    B_0(t) & \simeq 
    -\exp\left[ -\frac{\Omega_L^2+\Omega_S^2}{8\sigma^2}\right] \left(\frac{\Omega_S\Omega_L\cos\delta t}{4\sigma^2}\right).
\end{align}
The condition for optimum LO amplitude can be obtained from $\frac{dB_0}{d\Omega_L}=0$, to get (see Supplementary Information for the derivation)

\begin{equation}\label{eq:Optimum_Operating_point}
    \left[\frac{\Omega_L}{\sigma}\right]_{\mathrm{opt}} \simeq 2, 
\end{equation}

As the instantaneous Rabi rate in Eq.~\ref{eq:Instaneous_Rabi} is symmetric in $\Omega_S$ and $\Omega_L$, the nonlinear response is likewise symmetric with respect to interchange of the signal and LO amplitudes. However, when the two fields have comparable amplitudes, the optimum shifts from the weak-signal value of $\Omega_L/\sigma=2$. For the symmetric case $\Omega_S=\Omega_L$, the maximum fundamental response for a single tone excitation occurs at $\Omega_S/\sigma=\Omega_L/\sigma\simeq2.49$. 

Considering the practical implementation of simple two-photon Rydberg-atom receivers, the probe laser can be frequency locked reliably using saturated-absorption spectroscopy. However, locking the coupling laser exactly to zero detuning may not always be experimentally feasible. To examine the effect of such an offset, the measured and modeled responses for a coupling-laser detuning of $-10~\mathrm{MHz}$ are presented in Fig.~\ref{fig:Beatnote_differenatiation_2D_detuned}. The panel definitions and plotted quantities are identical to those in Fig.~\ref{fig:Fundamental_Harmonics_2D}. These results suggest that changes in the optical operating point can modify the nonlinear RF response of the Rydberg-atom receiver. However, under the detuned condition considered here, no overall improvement is observed in either the fundamental response or the suppression of the intermodulation components.  Even in this case, the overall structures of the intermodulation products show good agreement between the measured and modeled results. This further supports the validity of the proposed model for describing the nonlinear response under different optical operating conditions. A similar behavior is expected for other coupling-laser detunings, although the locations of the optimum RF operating points may shift with detuning.

\section{Effect of nonlinearity on EVM}
\label{sec:Nonlinear_EVM}

In continuation of the study of the nonlinear behavior of the Rydberg-atom receiver, we extend the analysis to the system-level performance metric, EVM \cite{hassun_effective_1997}, and relate it to the two-tone nonlinear response discussed above.

For this work, we utilize a QPSK modulation scheme. For the EVM measurements, a QPSK-modulated RF signal was generated using a vector signal generator and applied to the Rydberg-atom receiver. The resulting beat-note signal from the photodetector was measured using a vector signal analyzer. The received waveform was digitally demodulated, and the EVM was obtained from the recovered QPSK constellation. The measurements were repeated while varying the signal and LO RF amplitudes to determine the dependence of EVM on the operating point of the Rydberg-atom receiver.

By using the mathematical description of the response of the Rydberg-atom receiver in Eq.~\ref{eq:Beat_note_in_terms_of_Rabi_rate}, the EVM response is modeled. For the zero detuning case, $\Delta_C=0$, a numerical QPSK signal is generated and combined with the LO field to obtain the corresponding time-dependent Rabi frequency. The resulting waveform is then  digitally demodulated using the same filtering parameters as those used for the experimental data, and the EVM is calculated from the recovered QPSK symbols. This procedure is repeated for different signal and LO amplitudes to generate the modeled EVM response over the same operating range considered in the experiment.

The effect of the nonlinear response on the communication signal can be related to the nonlinear behavior observed in the two-tone measurements. For this purpose, the SNDR is extracted for both experimental data and modeled data from the two tone data shown in Fig.~\ref{fig:Fundamental_Harmonics_2D}. Here, the intermodulation products IM3, IM5, and IM7 are treated as distortion components, and their powers are combined with the noise floor to obtain the SNDR ratio over the two-dimensional space of signal and LO amplitude sweeps.  The fundamental component is taken as the desired signal power, while the combined contributions from the intermodulation products and the noise floor form the denominator.
\begin{equation}
    \mathrm{SNDR_{\mathrm{Two Tone}}} = \frac{P_{\mathrm{fund}}}{P_N+P_{\mathrm{IM3}}+P_{\mathrm{IM5}}+P_{\mathrm{IM7}}}. 
\end{equation}

\begin{figure}
    \centering
    \includegraphics[width=\linewidth]{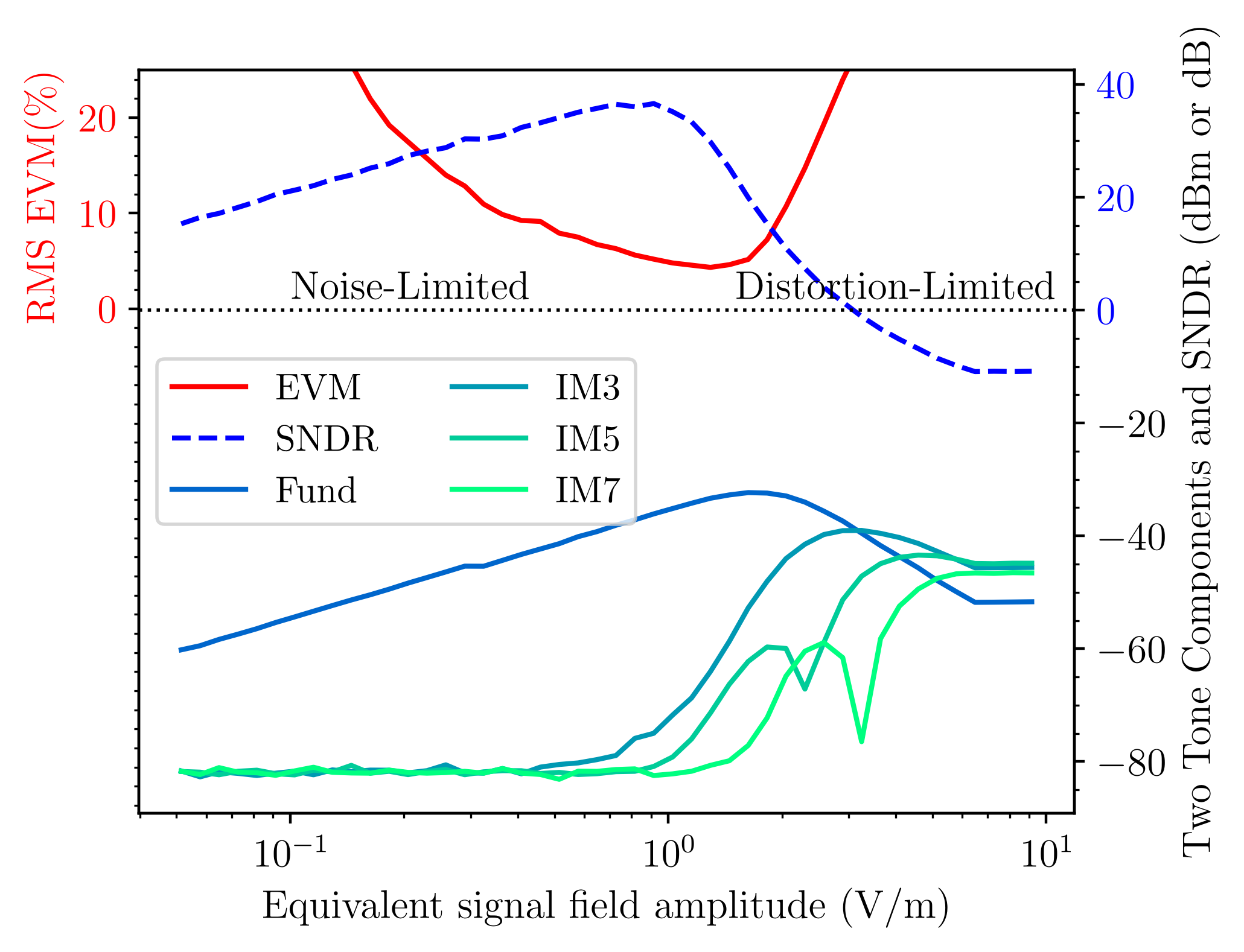}
    \caption{Noise-limited and distortion-limited regimes at a fixed LO amplitude of $\Omega_L/\sigma=5.10$. The RMS EVM corresponds to a QPSK bit rate of $50~\mathrm{kbps}$. The fundamental and intermodulation components extracted from the two-tone measurements are plotted together with the corresponding SNDR on the secondary $y$-axis. }
    \label{fig:EVM_SNDR_Summary_1D}
\end{figure}

Measured EVM maps for bit rates of $50~\mathrm{kbps}$ and $100~\mathrm{kbps}$ are shown in Fig.~\ref{fig:EVM_Experiment}(a) and (b), respectively. The corresponding modeled data is shown in Fig.~\ref{fig:EVM_Experiment}(d) and (e). For the purpose of presenting the data, the upper limit of the reported EVM is set to $25\%$ as it is considered the threshold above which reliable reconstruction of the QPSK signal becomes difficult \cite{schmogrow_error_2012}. To relate the two-tone behavior of the atomic receiver with the communication performance, the SNDR obtained from the two tone measurements is compared with the QPSK EVM. The corresponding SNDR values calculated from the two tone data are shown in Fig.~\ref{fig:EVM_Experiment}(c) and (f). All panels share the same axes and are presented in terms of both the calibrated RF electric-field amplitudes and the corresponding normalized RF Rabi frequencies. The IM3 null trajectory is overlaid on all the maps. 

The measured and modeled EVM maps show good qualitative agreement in their overall shapes and in the locations of the regions with increased EVM. A similar qualitative correspondence can also be observed between the EVM maps and the SNDR calculated from the two-tone measurements. Regions with reduced SNDR generally correspond to regions with increased EVM, showing that the nonlinear distortion observed in the two-tone response directly affects the quality of the digitally modulated signal.  Although the two-tone SNDR provides an indication of receiver nonlinearity, its optimum is not necessarily expected to coincide with the minimum EVM of a modulated waveform, since the resulting in-band distortion depends on the amplitude and phase statistics of the excitation even when the signals are normalized to the same average power \cite{carvalho_multisine_2008, gharaibeh_-band_2006} . Hence, the maximum two-tone SNDR can identify the low-distortion operating region, but the optimum operating point can shift for a root-raised-cosine (RRC) filtered QPSK waveform.

It can also be observed that, as the transmitted bit rate increases from $50~\mathrm{kbps}$ to $100~\mathrm{kbps}$, the usable region in the two-dimensional parameter space of $(\Omega_S,\Omega_L)$ becomes smaller. This indicates that increasing the data rate places a more stringent requirement on the operating point of the Rydberg-atom receiver in order to maintain an acceptable EVM.

Two regions of significant performance degradation in the parameter space of $(\Omega_S,\Omega_L)$ are labeled as zone A and zone B in Fig.~\ref{fig:EVM_Experiment}(a). In both zones, the high LO or signal amplitude drives the AT-split EIT spectral peaks away from the coupling-laser locking point, resulting in a reduced beat-note amplitude and consequently an increased EVM. Additionally, in zone B, the high signal amplitude results in increased intermodulation products as shown in Fig.~\ref{fig:Fundamental_Harmonics_2D}, which reduce the SNDR and consequently increase the EVM. This agreement between the two-tone characterization, the modeled EVM, and the measured EVM supports the role of the nonlinear atomic response in determining the communication performance of the Rydberg-atom receiver. It also indicates that the nonlinear response imposes practical limits on the range of signal and LO amplitudes over which the receiver can operate with an acceptable EVM.

 To clearly show the noise-limited and distortion-limited regimes, the SNDR obtained from the two-tone measurements is compared with the QPSK EVM at a fixed LO amplitude of $\Omega_L/\sigma=5.10$ in Fig.~\ref{fig:EVM_SNDR_Summary_1D}. The two-tone SNDR shows the inverse qualitative dependence on signal amplitude as the QPSK EVM. At low signal amplitudes, the receiver operates in the noise-limited regime, where the detected signal is small compared with the integrated noise and the EVM decreases as the signal amplitude is increased. As the signal amplitude is increased further, the nonlinear intermodulation products become significant, causing the SNDR to decrease and the EVM to increase. This marks the transition to the distortion-limited regime.

The minimum in the QPSK EVM occurs close to the maximum of the two-tone SNDR, although an exact alignment is not expected. The two-tone measurement probes the nonlinear response using two discrete frequencies, whereas the QPSK waveform occupies a continuous modulation bandwidth and contains a time-dependent amplitude and phase distribution. Consequently, the nonlinear products generated by the QPSK waveform are not identical to those obtained under two-tone excitation. In addition, the EVM includes the combined effects of noise, nonlinear amplitude and phase distortion, and the complete signal spectrum after matched filtering. Nevertheless, the close correspondence between the two measurements indicates that the nonlinear mixing behavior identified from the two-tone characterization captures the transition from noise-limited to distortion-limited operation of the Rydberg-atom receiver.

\section{Communication bandwidth limits}
\label{sec:Bandwidth}
 The achievable communication bandwidth of the Rydberg-atom receiver is determined not only by the intrinsic dynamical response of the atomic system, but also by the available SNDR. Rydberg-atom receivers can be limited by probe-photon shot noise associated with the optical detection process \cite{tang_noise_2025, bussey_quantum_2022, peng_signal-dependent_2026}. In a practical receiver, the measured noise floor can additionally contain contributions from detector electronic noise, laser intensity noise, and other technical noise sources. Together, these noise contributions determine the available receiver noise floor, while the nonlinear response introduces additional intermodulation distortion.  Therefore, even when the intrinsic dynamical response of the atomic system supports relatively large instantaneous bandwidth, the usable communication bandwidth can be smaller because increasing the signal bandwidth increases the integrated noise power, while increasing the signal amplitude is eventually limited by nonlinear distortion.
 
 As derived in the Supplementary Information, for an equivalent receiver noise bandwidth $B_n$, the RMS EVM can be expressed approximately as 
\begin{equation}
    \mathrm{EVM_{rms}^2} \simeq  \frac{N_0B_n}{P_S}+\frac{P_D}{P_S} ,
\end{equation}
where $N_0$ is the noise spectral density, $P_S$ is the desired signal power, and $P_D$ is the total in-band nonlinear distortion power. Here, we assume that the desired signal, nonlinear distortion, and noise contributions are mutually uncorrelated, and that the noise and nonlinear distortion can be treated as additive contributions to the received signal. The corresponding bandwidth available for a specified EVM requirement therefore has an upper limit specified as

\begin{equation}
    B_n \le \left[ \mathrm{EVM_{req}^2 - \frac{P_D}{P_S}} \right] \frac{P_S}{N_0}.
\end{equation}

This relation provides a direct connection between the nonlinear response of the Rydberg mixer and its achievable communication bandwidth. In the noise-limited regime $P_D\ll P_N$, the allowable bandwidth increases approximately linearly with the received signal power and decreases with increasing noise spectral density. Thus, for a fixed noise floor, a strong beat-note signal allows a large receiver bandwidth for a given EVM requirement.  However, this scaling is limited by the nonlinear distortion associated with increased RF signal or LO amplitudes as shown in Fig.~\ref{fig:Fundamental_Harmonics_2D}.

The increase in nonlinear distortion $P_D/P_S$ effectively reduces the achievable bandwidth. In the limiting case of 
\begin{equation}
    \frac{P_D}{P_S} > \mathrm{EVM_{req}^2},
\end{equation}
the required EVM cannot be achieved even in the limit of the vanishing noise bandwidth indicating a distortion-limited regime.

At low signal levels, the system can be noise limited, while at higher RF amplitudes it becomes distortion limited, even when the atomic response itself can support a larger instantaneous bandwidth. To isolate these effects from limitations associated with the finite dynamical response of the atomic system, we assume that the instantaneous bandwidth of the receiver exceeds $20~\mathrm{MHz}$ and that its frequency response is flat over the signal bandwidth. Under this assumption, the communication performance of the Rydberg-atom receiver with the nonlinear response discussed in the previous section is evaluated numerically. For this, random QPSK symbols are pulse-shaped using an RRC filter with roll-off factor of $\beta=0.35$ and modulated onto a $10~\mathrm{MHz}$ intermediate frequency carrier. The total RF coupling is represented by the coherent combination of the LO and modulated signal fields, and the resulting instantaneous Rabi amplitude is applied to the nonlinear Gaussian atomic response shown in Eq.~\ref{eq:Beat_note_in_terms_of_Rabi_rate} at zero coupling detuning ($\Delta_C=0$). The model is evaluated in the two-dimensional space of the normalized signal and LO Rabi amplitudes $(\Omega_L/\sigma,\Omega_S/\sigma)$. The receiver output is matched-filtered and sampled at the symbol rate, after which a best-fit complex gain is removed before evaluating the RMS EVM.  

The noise floor is set to  the measured noise floor of $-76.5~\mathrm{dBm}$ in  a $1~\mathrm{kHz}$ resolution bandwidth (RBW). The maximum measured beatnote power of $-30.66~\mathrm{dBm}$ is used to relate the normalized atomic response to the experimentally measured output power. The corresponding noise spectral density is converted into an equivalent time-domain noise amplitude and added to the receiver output after the nonlinear response calculation. Thus, the calculated EVM includes the combined  effects of detection noise and nonlinear distortion. This calculation is repeated for increasing data rates to determine how the communication performance evolves as the occupied signal bandwidth increases. The complete results in the space of $(\Omega_L/\sigma,\Omega_S/\sigma)$ are presented in the Supplementary Information.

\begin{figure}
    \centering
    \includegraphics[width=\linewidth]{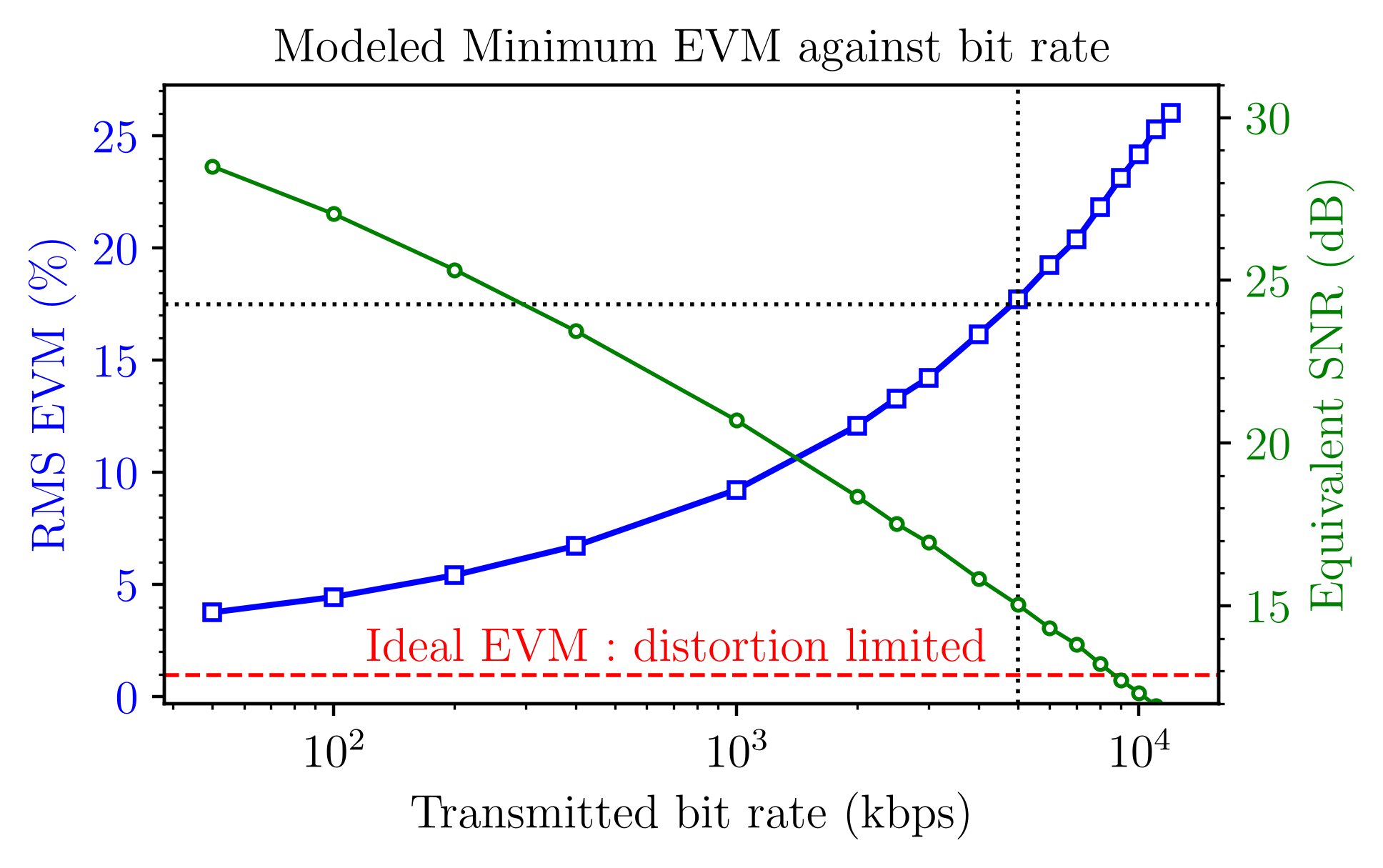}
    \caption{The EVM is extracted from the model for various bit rates, assuming that the system bandwidth exceeds the IF frequency of $10~\mathrm{MHz}$ and frequency response is assumed flat over the occupied signal bandwidth. The corresponding equivalent SNR is also plotted for comparison. For a required EVM of $17.5\%$, the maximum achievable bit rate is approximately $5~\mathrm{Mbps}$. In the ideal case, where the noise contribution is neglected, the minimum achievable EVM is limited by the nonlinear distortion and is indicated by the red line.}
    \label{fig:EVM_SNDR_Model}
\end{figure}

From the modeled data, the minimum EVM obtained at each data rate is plotted in Fig~\ref{fig:EVM_SNDR_Model}, together with the corresponding equivalent SNR, defined as
$\mathrm{SNR_{dB}^{eq}}=-20\log_{10}\left(\mathrm{EVM_{rms}}\right)$ \cite{mahmoud_error_2009}. The minimum achievable EVM increases with data rate, with a corresponding decrease in the equivalent SNR, thereby reducing the range of $(\Omega_L/\sigma,\Omega_S/\sigma)$ over which the required EVM can be maintained. For the experimental system considered in this work, characterized by $\sigma=8.97~\mathrm{MHz}$ and a measured photodetection noise floor of $-76.5~\mathrm{dBm}$, the model predicts a maximum data rate of approximately $5~\mathrm{Mbps}$ for an EVM limit of $17.5\%$. Here, $17.5\%$ is used as a reference value for QPSK based on the EVM requirement specified in the 3GPP standards \cite{3gpp_specification}.

From the data in Fig.~\ref{fig:Fundamental_Harmonics_2D}, the maximum SNR is $\approx50~\mathrm{dB}$; however, when the intermodulation distortion is included, the maximum SNDR is reduced to $\approx40~\mathrm{dB}$.

\section{Discussion}
\label{sec:Discussion}
 
Under typical experimental conditions considered here, the EIT lineshape is well-approximated by a Gaussian function. By extracting a single characteristic parameter, namely the standard deviation $\sigma$ of the Gaussian EIT resonance, the nonlinear heterodyne response of the receiver can be modeled. When the signal and LO Rabi frequencies are normalized as $\Omega_s/\sigma$  and $\Omega_L/\sigma$, respectively, the model captures the receiver response over a broad range of signal and LO amplitudes and shows good agreement with the experimental results.

In conventional RF mixers, suppression of the IM3 component is commonly achieved using engineered linearization techniques, including transconductance cancellation \cite{liang_new_2008}, nonlinear feedback \cite{amirabadi_high_2011, seyedi_uwb_2021}, and feedforward cancellation \cite{li_linearization_2019}. These approaches generally introduce an additional nonlinear pathway whose third-order response partially cancels the IM3 generated by the primary mixing process. In contrast, the Rydberg-atom mixer considered here exhibits an intrinsic IM3 cancellation mechanism arising directly from the nonlinear response associated  with the EIT lineshape, without requiring an additional linearization path. 

In particular, the model captures the evolution of the odd-order intermodulation products that limit the linearity of the receiver. Analytical conditions for the cancellation of selected intermodulation products are obtained for the Gaussian response and agree well with the experimentally observed minima. However, the null trajectories corresponding to different intermodulation orders do not generally coincide. Consequently, there is no single operating condition at which all the intermodulation products considered here are simultaneously minimized. The optimum operating region is therefore determined by the dominant distortion product (IM3) along with the corresponding heterodyne conversion efficiency.

The intermodulation response also has direct implications for the achievable  communication bandwidth by the Rydberg-atom receiver. As discussed in the Supplementary Information, the EVM can be separated approximately into contributions from noise and nonlinear distortion. For a given noise spectral density, the allowable receiver noise bandwidth is dictated by the in-band distortion components. This EVM-limited bandwidth needs to be distinguished from the intrinsic dynamical bandwidth of the atomic receiver. The practically usable bandwidth is ultimately limited by both the nonlinear distortion plus photodetection noise and the frequency-dependent amplitude and phase response of the atomic system.

The Gaussian approximation also introduces several limitations. First, representing the EIT resonance by a Gaussian is itself an approximation, and is only valid when inhomogeneous broadening mechanisms broaden the lineshape past the Doppler residual limited lineshape \cite{schlossberger_fundamental_2026,holloway_autlertownes_2026}. Second, the  effective spectral linewidth can change at high-rf amplitudes as the atomic response becomes strongly dressed, making the assumption of a constant $\sigma$ increasingly inaccurate. The present model also neglects the frequency-dependent response of the atomic receiver and the photodetection chain, including finite receiver bandwidth, frequency-dependent conversion gain, and phase-delay variations. These effects are expected to become increasingly important for broadband and high-data-rate modulations.

Despite these limitations, the Gaussian model provides a simple analytical framework for interpreting the nonlinear operating regimes of Rydberg-atom heterodyne receivers. Expressing the response in terms of the normalized parameters $\Omega_s/\sigma$ and $\Omega_L/\sigma$ enables identification of operating regions that balance heterodyne conversion efficiency, intermodulation distortion, noise, and the resulting EVM. The model also provides a direct connection between the atomic spectral response and the communication-level receiver metrics without requiring full density matrix calculation for every operating point.

Future work can extend this approach to different Rydberg states and linewidths, nonzero optical detunings, different excitation and polarization conditions, and alternative modulation formats. Incorporating the full frequency-dependent atomic response, phase noise, and nonlinear phase distortion would further enable optimization of  Rydberg-atom receivers for broadband communication applications. 

In this work, we investigated the nonlinear response of a Rydberg-atom heterodyne receiver and established a direct connection between the atomic EIT lineshape and the generation of intermodulation distortion. We showed that, over the operating regime considered here, the nonlinear mixing response can be described by using a Gaussian model parametrized primarily by the EIT Gaussian lineshape standard deviation $\sigma$. The model reproduces the experimentally observed behavior of the fundamental beat note and higher-order intermodulation products and yields analytical conditions under which the selected nonlinear products are suppressed. As these conditions for different intermodulation products do not coincide, the optimum receiver operating point is a balance between the heterodyne conversion efficiency, nonlinear distortion, and the noise from photodetection. 

The nonlinear response also establishes a direct link between atomic physics and communication-level performance. In-band intermodulation products contribute to the residual error together with the detector noise and therefore limit the achievable EVM and receiver bandwidth. The results presented in this work identify the EIT spectral lineshape parameter ($\sigma$) and the relative signal and LO Rabi frequencies as the key parameters governing not only the nonlinear spectrum but also the usable operating region of the atomic receiver. These results provide a framework for treating the Rydberg-atom receiver as a nonlinear RF mixer rather than solely as an electric-field sensor. It provides the ability to predict and control nonlinear mixing products providing a route toward optimizing Rydberg receivers for high-dynamic range electrometry and digitally modulated communication signals.

\section{Acknowledgments}
The authors thank Joshua M. Kast for the initial discussions and Bryan T. Bosworth for kindly providing the hardware support. This research  was supported by NIST under the NIST-on-a-Chip program. A contribution of the U.S. government, this work is not subject to copyright in the United States. The authors declare no conflict of interest.

\section{Data Availability}
All of the data presented and used to
support the conclusions of this article are available at \cite{talashila_data_nodate}.

\section{Code Availability}
The code used for the analysis and  to
support the conclusions of this article is available at \cite{talashila_data_nodate}.

\section*{Author Contributions}
R.T. conceived and designed the study, led the experimental investigations, developed the theoretical framework and analytical models, performed numerical simulations, analyzed and interpreted the data, and prepared the manuscript. S.B.O. contributed to the experimental work. N.S. contributed to scientific discussions. C.L.H. supervised the research and contributed to scientific discussions. N.P. supervised the research and contributed to scientific discussions and guided the presentation of the results.  All authors contributed to the writing and revision of the manuscript and approved its final version.

\section*{Competing Interests}
The authors declare no competing interests.

\bibliography{Refs}

@article{goncalves_nonlinearities_2026,
    title = {Nonlinearities in atomic radio receivers: {Harmonic} and intermodulation distortion},
    volume = {1},
    shorttitle = {Nonlinearities in atomic radio receivers},
    url = {https://link.aps.org/doi/10.1103/3pht-dvg9},
    doi = {10.1103/3pht-dvg9},
    urldate = {2026-08-20},
    journal = {APS Open Science},
    publisher = {American Physical Society},
    author = {Gonçalves, Luís Felipe and Zhang, Teng and Raithel, Georg and Anderson, David A.},
    month = aug,
    year = {2026},
    pages = {000083},
}

@article{jing_atomic_2020,
    title = {Atomic superheterodyne receiver based on microwave-dressed {Rydberg} spectroscopy},
    volume = {16},
    copyright = {2020 The Author(s), under exclusive licence to Springer Nature Limited},
    issn = {1745-2481},
    url = {https://www.nature.com/articles/s41567-020-0918-5},
    doi = {10.1038/s41567-020-0918-5},
    number = {9},
    urldate = {2024-09-10},
    journal = {Nature Physics},
    publisher = {Nature Publishing Group},
    author = {Jing, Mingyong and Hu, Ying and Ma, Jie and Zhang, Hao and Zhang, Linjie and Xiao, Liantuan and Jia, Suotang},
    month = sep,
    year = {2020},
    pages = {911--915},
}

@article{simons_rydberg_2019,
    title = {A {Rydberg} atom-based mixer: {Measuring} the phase of a radio frequency wave},
    volume = {114},
    issn = {0003-6951, 1077-3118},
    shorttitle = {A {Rydberg} atom-based mixer},
    url = {https://pubs.aip.org/apl/article/114/11/114101/36735/A-Rydberg-atom-based-mixer-Measuring-the-phase-of},
    doi = {10.1063/1.5088821},
    number = {11},
    urldate = {2024-08-13},
    journal = {Applied Physics Letters},
    author = {Simons, Matthew T. and Haddab, Abdulaziz H. and Gordon, Joshua A. and Holloway, Christopher L.},
    month = mar,
    year = {2019},
    pages = {114101},
}

@article{holloway_broadband_2014,
    title = {Broadband {Rydberg} {Atom}-{Based} {Electric}-{Field} {Probe} for {SI}-{Traceable}, {Self}-{Calibrated} {Measurements}},
    volume = {62},
    issn = {1558-2221},
    url = {https://ieeexplore.ieee.org/document/6910267},
    doi = {10.1109/TAP.2014.2360208},
    number = {12},
    urldate = {2024-09-11},
    journal = {IEEE Transactions on Antennas and Propagation},
    author = {Holloway, Christopher L. and Gordon, Joshua A. and Jefferts, Steven and Schwarzkopf, Andrew and Anderson, David A. and Miller, Stephanie A. and Thaicharoen, Nithiwadee and Raithel, Georg},
    month = dec,
    year = {2014},
    note = {Conference Name: IEEE Transactions on Antennas and Propagation},
    pages = {6169--6182},
}

@article{sedlacek_microwave_2012,
    title = {Microwave electrometry with {Rydberg} atoms in a vapour cell using bright atomic resonances},
    volume = {8},
    copyright = {2012 Springer Nature Limited},
    issn = {1745-2481},
    url = {https://www.nature.com/articles/nphys2423},
    doi = {10.1038/nphys2423},
    number = {11},
    urldate = {2024-09-11},
    journal = {Nature Physics},
    publisher = {Nature Publishing Group},
    author = {Sedlacek, Jonathon A. and Schwettmann, Arne and Kübler, Harald and Löw, Robert and Pfau, Tilman and Shaffer, James P.},
    month = nov,
    year = {2012},
    pages = {819--824},
}

@article{meyer_assessment_2020,
    title = {Assessment of {Rydberg} atoms for wideband electric field sensing},
    volume = {53},
    issn = {0953-4075},
    url = {https://doi.org/10.1088/1361-6455/ab6051},
    doi = {10.1088/1361-6455/ab6051},
    number = {3},
    urldate = {2026-07-03},
    journal = {Journal of Physics B: Atomic, Molecular and Optical Physics},
    publisher = {IOP Publishing},
    author = {Meyer, David H and Castillo, Zachary A and Cox, Kevin C and Kunz, Paul D},
    month = jan,
    year = {2020},
    pages = {034001},
}

@article{liu_continuous-frequency_2022,
    title = {Continuous-{Frequency} {Microwave} {Heterodyne} {Detection} in an {Atomic} {Vapor} {Cell}},
    volume = {18},
    url = {https://link.aps.org/doi/10.1103/PhysRevApplied.18.054003},
    doi = {10.1103/PhysRevApplied.18.054003},
    number = {5},
    urldate = {2026-02-02},
    journal = {Physical Review Applied},
    publisher = {American Physical Society},
    author = {Liu, Xiao-Hong and Liao, Kai-Yu and Zhang, Zuan-Xian and Tu, Hai-Tao and Bian, Wu and Li, Zhong-Qi and Zheng, Shun-Yuan and Li, He-He and Huang, Wei and Yan, Hui and Zhu, Shi-Liang},
    month = nov,
    year = {2022},
    pages = {054003},
}

@article{shabanzadeh_circuit_2026,
    title = {Circuit {Linearization} {Methods} and {Theoretical} {Nonlinearity} {Analysis} in {Mixers}: {A} {Review}},
    volume = {14},
    issn = {2169-3536},
    shorttitle = {Circuit {Linearization} {Methods} and {Theoretical} {Nonlinearity} {Analysis} in {Mixers}},
    url = {https://ieeexplore.ieee.org/document/11657279/},
    doi = {10.1109/ACCESS.2026.3724295},
    urldate = {2026-09-08},
    journal = {IEEE Access},
    author = {Shabanzadeh, Negar and Pärssinen, Aarno and Rahkonen, Timo},
    year = {2026},
    pages = {130647--130668},
}

@article{jiao_spectroscopy_2016,
    title = {Spectroscopy of cesium {Rydberg} atoms in strong radio-frequency fields},
    volume = {94},
    url = {https://link.aps.org/doi/10.1103/PhysRevA.94.023832},
    doi = {10.1103/PhysRevA.94.023832},
    number = {2},
    urldate = {2026-08-18},
    journal = {Physical Review A},
    publisher = {American Physical Society},
    author = {Jiao, Yuechun and Han, Xiaoxuan and Yang, Zhiwei and Li, Jingkui and Raithel, Georg and Zhao, Jianming and Jia, Suotang},
    month = aug,
    year = {2016},
    pages = {023832},
}

@article{hao_rydberg_2019,
    title = {Rydberg electromagnetically induced transparency and {Autler}–{Townes} splitting in a weak radio-frequency electric field*},
    volume = {28},
    issn = {1674-1056},
    url = {https://doi.org/10.1088/1674-1056/28/5/053202},
    doi = {10.1088/1674-1056/28/5/053202},
    number = {5},
    urldate = {2026-09-17},
    journal = {Chinese Physics B},
    publisher = {Chinese Physical Society and IOP Publishing Ltd},
    author = {Hao, Liping and Xue, Yongmei and Fan, Jiabei and Jiao, Yuechun and Zhao, Jianming and Jia, Suotang},
    month = may,
    year = {2019},
    pages = {053202},
}

@article{hao_nonlinearity_2019,
    title = {Nonlinearity of {Microwave} {Electric} {Field} {Coupled} {Rydberg} {Electromagnetically} {Induced} {Transparency} and {Autler}-{Townes} {Splitting}},
    volume = {9},
    copyright = {http://creativecommons.org/licenses/by/3.0/},
    issn = {2076-3417},
    url = {https://www.mdpi.com/2076-3417/9/8/1720},
    doi = {10.3390/app9081720},
    number = {8},
    urldate = {2026-09-17},
    journal = {Applied Sciences},
    publisher = {Multidisciplinary Digital Publishing Institute},
    author = {Hao, Liping and Xue, Yongmei and Fan, Jiabei and Jiao, Yuechun and Zhao, Jianming and Jia, Suotang},
    month = jan,
    year = {2019},
    pages = {1720},
}

@article{holloway_electric_2017,
    title = {Electric field metrology for {SI} traceability: {Systematic} measurement uncertainties in electromagnetically induced transparency in atomic vapor},
    volume = {121},
    issn = {0021-8979, 1089-7550},
    shorttitle = {Electric field metrology for {SI} traceability},
    url = {https://pubs.aip.org/jap/article/121/23/233106/143771/Electric-field-metrology-for-SI-traceability},
    doi = {10.1063/1.4984201},
    number = {23},
    urldate = {2024-09-11},
    journal = {Journal of Applied Physics},
    author = {Holloway, Christopher L. and Simons, Matt T. and Gordon, Joshua A. and Dienstfrey, Andrew and Anderson, David A. and Raithel, Georg},
    month = jun,
    year = {2017},
    pages = {233106},
}

@article{mahmoud_error_2009,
    title = {Error vector magnitude to {SNR} conversion for nondata-aided receivers},
    volume = {8},
    issn = {1558-2248},
    url = {https://ieeexplore.ieee.org/document/4927484},
    doi = {10.1109/TWC.2009.080862},
    number = {5},
    urldate = {2026-09-28},
    journal = {IEEE Transactions on Wireless Communications},
    author = {Mahmoud, Hisham A. and Arslan, Huseyin},
    month = may,
    year = {2009},
    pages = {2694--2704},
}

@article{sibalic_arc_2017,
    title = {{ARC}: {An} open-source library for calculating properties of alkali {Rydberg} atoms},
    volume = {220},
    issn = {0010-4655},
    shorttitle = {{ARC}},
    url = {https://www.sciencedirect.com/science/article/pii/S0010465517301972},
    doi = {10.1016/j.cpc.2017.06.015},
    urldate = {2025-11-13},
    journal = {Computer Physics Communications},
    author = {Šibalić, N. and Pritchard, J. D. and Adams, C. S. and Weatherill, K. J.},
    month = nov,
    year = {2017},
    pages = {319--331},
}

@article{tang_noise_2025,
    title = {Noise performance of a {Rydberg} atomic receiver},
    volume = {112},
    url = {https://link.aps.org/doi/10.1103/3zlx-g5xl},
    doi = {10.1103/3zlx-g5xl},
    number = {4},
    urldate = {2026-09-29},
    journal = {Physical Review A},
    publisher = {American Physical Society},
    author = {Tang, Yu and Zhou, Hanbin and Yang, Chuang and Ren, Shuang and Wang, Siyuan and Lu, Chenxi},
    month = oct,
    year = {2025},
    pages = {043106},
}

@article{bussey_quantum_2022,
    title = {Quantum {Shot} {Noise} {Limit} in a {Rydberg} {RF} {Receiver} {Compared} to {Thermal} {Noise} {Limit} in a {Conventional} {Receiver}},
    volume = {6},
    issn = {2475-1472},
    url = {https://ieeexplore.ieee.org/document/9873938},
    doi = {10.1109/LSENS.2022.3203465},
    number = {9},
    urldate = {2026-09-29},
    journal = {IEEE Sensors Letters},
    author = {Bussey, Liam W. and Burton, Fraser A. and Bongs, Kai and Goldwin, Jonathan and Whitley, Tim},
    month = sep,
    year = {2022},
    pages = {1--4},
}

@misc{peng_signal-dependent_2026,
    title = {Signal-{Dependent} {Shot} {Noise} {Modeling} of {Rydberg} {Atomic} {Quantum} {Receivers}: {A} {Design} {Perspective}},
    shorttitle = {Signal-{Dependent} {Shot} {Noise} {Modeling} of {Rydberg} {Atomic} {Quantum} {Receivers}},
    url = {http://arxiv.org/abs/2605.10350},
    doi = {10.48550/arXiv.2605.10350},
    urldate = {2026-09-29},
    publisher = {arXiv},
    author = {Peng, Qihao and Luo, Qu and Gong, Tierui and Ye, Neng and Wu, Jizhou and Pan, Cunhua and Elkashlan, Maged and Xiao, Pei and Yuen, Chau and Karagiannidis, George K. and Wang, Jiangzhou},
    month = may,
    year = {2026},
    note = {arXiv:2605.10350 [eess.SP]},
}

@article{schmogrow_error_2012,
    title = {Error {Vector} {Magnitude} as a {Performance} {Measure} for {Advanced} {Modulation} {Formats}},
    volume = {24},
    issn = {1941-0174},
    url = {https://ieeexplore.ieee.org/document/6046222},
    doi = {10.1109/LPT.2011.2172405},
    number = {1},
    urldate = {2026-09-30},
    journal = {IEEE Photonics Technology Letters},
    author = {Schmogrow, Rene and Nebendahl, Bernd and Winter, Marcus and Josten, Arne and Hillerkuss, David and Koenig, Swen and Meyer, Joachim and Dreschmann, Michael and Huebner, Michael and Koos, Christian and Becker, Juergen and Freude, Wolfgang and Leuthold, Juerg},
    month = jan,
    year = {2012},
    pages = {61--63},
}

@article{holloway_autlertownes_2026,
    title = {Autler–{Townes} splitting in {Rydberg} atoms: transition dipole matrix element extraction and field efficiency analysis},
    volume = {59},
    issn = {0953-4075},
    shorttitle = {Autler–{Townes} splitting in {Rydberg} atoms},
    url = {https://doi.org/10.1088/1361-6455/ae407f},
    doi = {10.1088/1361-6455/ae407f},
    number = {3},
    urldate = {2026-10-05},
    journal = {Journal of Physics B: Atomic, Molecular and Optical Physics},
    publisher = {IOP Publishing},
    author = {Holloway, Brian C and Chase, Gavin M and Harrell, Lee E and Hutchinson, David W and North, William K and Catapano, Tyler M and Custodio, Christian F and Kirschenbaum, Simon L and Speer, Michael C and Kashinski, David O},
    month = feb,
    year = {2026},
    pages = {035002},
}

@book{pedro_intermodulation_2003,
    address = {Boston},
    series = {Artech {House} microwave library},
    title = {Intermodulation distortion in microwave and wireless circuits},
    isbn = {978-1-58053-356-0},
    publisher = {Artech House},
    author = {Pedro, José Carlos and Carvalho, Nuno Borges},
    year = {2003},
}

@article{lui_passive_1990,
    title = {Passive intermodulation interference in communication systems},
    volume = {2},
    issn = {0954-0695, 2051-218X},
    url = {http://digital-library.theiet.org/doi/10.1049/ecej%3A19900029},
    doi = {10.1049/ecej:19900029},
    number = {3},
    urldate = {2026-10-05},
    journal = {Electronics \& Communication Engineering Journal},
    author = {Lui, P.L.},
    month = jun,
    year = {1990},
    pages = {109--118},
}

@article{liang_new_2008,
    title = {A {New} {Linearization} {Technique} for {CMOS} {RF} {Mixer} {Using} {Third}-{Order} {Transconductance} {Cancellation}},
    volume = {18},
    issn = {1558-1764},
    url = {https://ieeexplore.ieee.org/document/4497814/},
    doi = {10.1109/LMWC.2008.922129},
    number = {5},
    urldate = {2026-10-09},
    journal = {IEEE Microwave and Wireless Components Letters},
    author = {Liang, Kung-Hao and Lin, Chi-Hsein and Chang, Hong-Yeh and Chan, Yi-Jen},
    month = may,
    year = {2008},
    pages = {350--352},
}

@article{amirabadi_high_2011,
    title = {High {IIP3} and {Low}-{Noise} {CMOS} {Mixer} {Using} {Non}-linear {Feedback} {Technique}},
    volume = {30},
    issn = {1531-5878},
    url = {https://doi.org/10.1007/s00034-011-9301-1},
    doi = {10.1007/s00034-011-9301-1},
    number = {4},
    urldate = {2026-10-09},
    journal = {Circuits, Systems, and Signal Processing},
    author = {Amirabadi, Amir and Chehelcheraghi, Mojtaba and Kamarei, Mahmoud},
    month = aug,
    year = {2011},
    pages = {721--739},
}

@article{li_linearization_2019,
    title = {Linearization of {Active} {Downconversion} {Mixers} at the {IF} {Using} {Feedforward} {Cancellation}},
    volume = {66},
    issn = {1558-0806},
    url = {https://ieeexplore-ieee-org.colorado.idm.oclc.org/document/8574959},
    doi = {10.1109/TCSI.2018.2883920},
    number = {4},
    urldate = {2026-10-09},
    journal = {IEEE Transactions on Circuits and Systems I: Regular Papers},
    author = {Li, Hao and Saavedra, Carlos E.},
    month = apr,
    year = {2019},
    pages = {1620--1631},
}

@article{seyedi_uwb_2021,
    title = {{UWB} down-conversion mixer using an {IM3} cancellation modified technique for zero and low {IF} applications},
    volume = {109},
    issn = {1879-2391},
    url = {https://www.sciencedirect.com/science/article/pii/S0026269220305826},
    doi = {10.1016/j.mejo.2020.104983},
    urldate = {2026-10-09},
    journal = {Microelectronics Journal},
    author = {Seyedi, Hosein and Dehdasht-Heydari, Ramin and Roshani, Saeed},
    month = mar,
    year = {2021},
    pages = {104983},
}

@article{schlossberger_fundamental_2026,
  title = {Fundamental linewidth limit of electromagnetically induced transparency in a thermal {Rydberg} ladder},
  volume = {8},
  url = {https://link.aps.org/doi/10.1103/g2z2-2w2h},
  doi = {10.1103/g2z2-2w2h},
  number = {3},
  urldate = {2026-10-23},
  journal = {Physical Review Research},
  publisher = {American Physical Society},
  author = {Schlossberger, Noah and Prajapati, Nikunjkumar and Artusio-Glimpse, Alexandra B. and Berweger, Samuel and Holloway, Christopher L.},
  month = sep,
  year = {2026},
  pages = {L032046},
}

@techreport{3gpp_specification,
    author      = {{3GPP}},
    title       = {{NR; Base Station (BS) radio transmission and reception}},
    type        = {TS},
    number      = {38.104},
    version     = {19.6.0},
    institution = {3GPP},
    series      = {Release 19},
    year        = {2026},
    month       = {September},
    url         = {https://www.3gpp.org/dynareport/38104.htm}
}

@article{yang_highly_2024,
  title = {Highly sensitive microwave electrometry with enhanced instantaneous bandwidth},
  volume = {21},
  url = {https://link.aps.org/doi/10.1103/PhysRevApplied.21.L031003},
  doi = {10.1103/PhysRevApplied.21.L031003},
  number = {3},
  urldate = {2026-11-12},
  journal = {Physical Review Applied},
  publisher = {American Physical Society},
  author = {Yang, Bowen and Yan, Yuhan and Li, Xuejie and Xiao, Ling and Li, Xiaolin and Chen, L.Q. and Deng, Jianliao and Cheng, Huadong},
  month = mar,
  year = {2024},
  pages = {L031003},
}

@article{jiao_-demand_2026,
  title = {On-demand instantaneous bandwidth microwave receiver via scalable {Rydberg} {Vapor} cell array with {Stark} comb},
  copyright = {2026 The Author(s)},
  issn = {2399-3650},
  url = {https://www.nature.com/articles/s42005-026-02941-9},
  doi = {10.1038/s42005-026-02941-9},
  urldate = {2026-11-12},
  journal = {Communications Physics},
  publisher = {Nature Publishing Group},
  author = {Jiao, Yuechun and Yin, Yuwen and He, Yunhui and Hu, Jinlian and Lu, Cheng and Bai, Jingxu and Bai, Zhengyang and Li, Weibin and Jia, Suotang and Zhao, Jianming},
  month = nov,
  year = {2026},
}

@article{cox_quantum-limited_2018,
  title = {Quantum-{Limited} {Atomic} {Receiver} in the {Electrically} {Small} {Regime}},
  volume = {121},
  url = {https://link.aps.org/doi/10.1103/PhysRevLett.121.110502},
  doi = {10.1103/PhysRevLett.121.110502},
  number = {11},
  urldate = {2026-11-12},
  journal = {Physical Review Letters},
  publisher = {American Physical Society},
  author = {Cox, Kevin C. and Meyer, David H. and Fatemi, Fredrik K. and Kunz, Paul D.},
  month = sep,
  year = {2018},
  pages = {110502},
}

@article{manchaiah_probing_2026,
  title = {Probing bandwidth and sensitivity in {Rydberg} atom sensing via optical homodyne and rf heterodyne detection},
  volume = {113},
  url = {https://link.aps.org/doi/10.1103/jsbl-45t9},
  doi = {10.1103/jsbl-45t9},
  number = {1},
  urldate = {2026-11-12},
  journal = {Physical Review A},
  publisher = {American Physical Society},
  author = {Manchaiah, Dixith and Oliver, Stone and Berweger, Samuel and Holloway, Christopher L. and Prajapati, Nikunjkumar},
  month = jan,
  year = {2026},
  pages = {013729},
}

@article{otto_data_2021,
  title = {Data capacity scaling of a distributed {Rydberg} atomic receiver array},
  volume = {129},
  issn = {0021-8979},
  url = {https://doi.org/10.1063/5.0048415},
  doi = {10.1063/5.0048415},
  number = {15},
  urldate = {2026-11-12},
  journal = {Journal of Applied Physics},
  author = {Otto, J. Susanne and Hunter, Marisol K. and Kjærgaard, Niels and Deb, Amita B.},
  month = apr,
  year = {2021},
  pages = {154503},
}

@article{nowosielski_warm_2024,
  title = {Warm {Rydberg} atom-based quadrature amplitude-modulated receiver},
  volume = {32},
  issn = {1094-4087},
  url = {https://opg.optica.org/oe/abstract.cfm?uri=oe-32-17-30027},
  doi = {10.1364/OE.529977},
  number = {17},
  urldate = {2026-11-12},
  journal = {Optics Express},
  publisher = {Optica Publishing Group},
  author = {Nowosielski, Jan and Jastrzębski, Marcin and Halavach, Pavel and Łukanowski, Karol and Jarzyna, Marcin and Mazelanik, Mateusz and Wasilewski, Wojciech and Parniak, Michał},
  month = aug,
  year = {2024},
  pages = {30027--30039},
}

@article{qu_channel-capacity_2026,
  title = {Channel-capacity optimization in superheterodyne {Rydberg} atomic receivers},
  volume = {114},
  url = {https://link.aps.org/doi/10.1103/wxx5-1zxt},
  doi = {10.1103/wxx5-1zxt},
  number = {2},
  urldate = {2026-11-12},
  journal = {Physical Review A},
  publisher = {American Physical Society},
  author = {Qu, Chuan and Zhang, Jian},
  month = aug,
  year = {2026},
  pages = {023720},
}

@article{zhu_general_2026,
  title = {General {Signal} {Model} and {Capacity} {Limit} for {Rydberg} {Quantum} {Information} {System}},
  volume = {25},
  copyright = {https://ieeexplore.ieee.org/Xplorehelp/downloads/license-information/IEEE.html},
  issn = {1536-1276, 1558-2248},
  url = {https://ieeexplore.ieee.org/document/11278503/},
  doi = {10.1109/TWC.2025.3637029},
  urldate = {2026-11-12},
  journal = {IEEE Transactions on Wireless Communications},
  author = {Zhu, Jieao and Dai, Linglong},
  year = {2026},
  pages = {8292--8307},
}

@article{carvalho_multisine_2008,
  title = {Multisine signals for wireless system test and design [{Application} {Notes}]},
  volume = {9},
  issn = {1557-9581},
  url = {https://ieeexplore.ieee.org/document/4519501},
  doi = {10.1109/MMM.2008.919938},
  number = {3},
  urldate = {2026-11-20},
  journal = {IEEE Microwave Magazine},
  author = {Carvalho, Nuno B. and Remley, Kate A. and Schreurs, Dominique and Gard, Kevin G.},
  month = jun,
  year = {2008},
  pages = {122--138},
}

@article{gharaibeh_-band_2006,
  title = {In-band distortion of multisines},
  volume = {54},
  issn = {1557-9670},
  url = {https://ieeexplore.ieee.org/document/1668339},
  doi = {10.1109/TMTT.2006.879170},
  number = {8},
  urldate = {2026-11-20},
  journal = {IEEE Transactions on Microwave Theory and Techniques},
  author = {Gharaibeh, K.M. and Gard, K.G. and Steer, M.B.},
  month = aug,
  year = {2006},
  pages = {3227--3236},
}

@misc{talashila_characteristics_2026,
  title = {Characteristics of {VHF} and {UHF} band detection using {Stark} shifting in {Rydberg}-atom sensors},
  url = {http://arxiv.org/abs/2610.40952},
  doi = {10.48550/arXiv.2610.40952},
  publisher = {arXiv},
  author = {Talashila, Rajavardhan and Schlossberger, Noah and Mantia, David S. La and Halasi-Kun, Abby and Kayim, Baran and Viray, Michael A. and Wyllie, Robert and Holloway, Christopher L. and Prajapati, Nikunjkumar},
  month = nov,
  year = {2026},
  note = {arXiv:2610.40952 [physics.atom-ph]},
}

@article{hammerland_tuning_2026,
  title = {Tuning {Rydberg} sensor capabilities with quantum media},
  volume = {8},
  issn = {2639-0213},
  url = {https://doi.org/10.1116/5.0353346},
  doi = {10.1116/5.0353346},
  number = {4},
  urldate = {2026-11-21},
  journal = {AVS Quantum Science},
  author = {Hammerland, Daniel and Talashila, Rajavardhan and Manchaiah, Dixith and Schlossberger, Noah and Prajapati, Nikunjkumar and McKee, Erik and Highman, Michael A. and Simons, Matthew T. and Berweger, Samuel and Artusio-Glimpse, Alexandra B. and Holloway, Christopher L.},
  month = nov,
  year = {2026},
  pages = {044403},
}

@article{liu_highly_2022,
  title = {Highly {Sensitive} {Measurement} of a {Megahertz} rf {Electric} {Field} with a {Rydberg}-{Atom} {Sensor}},
  volume = {18},
  url = {https://link.aps.org/doi/10.1103/PhysRevApplied.18.014045},
  doi = {10.1103/PhysRevApplied.18.014045},
  number = {1},
  urldate = {2026-11-21},
  journal = {Physical Review Applied},
  publisher = {American Physical Society},
  author = {Liu, Bang and Zhang, Li-Hua and Liu, Zong-Kai and Zhang, Zheng-Yuan and Zhu, Zhi-Han and Gao, Wei and Guo, Guang-Can and Ding, Dong-Sheng and Shi, Bao-Sen},
  month = jul,
  year = {2022},
  pages = {014045},
}

@article{thaicharoen_electromagnetically_2019,
  title = {Electromagnetically induced transparency, absorption, and microwave-field sensing in a {Rb} vapor cell with a three-color all-infrared laser system},
  volume = {100},
  url = {https://link.aps.org/doi/10.1103/PhysRevA.100.063427},
  doi = {10.1103/PhysRevA.100.063427},
  number = {6},
  urldate = {2025-01-23},
  journal = {Physical Review A},
  publisher = {American Physical Society},
  author = {Thaicharoen, N. and Moore, K. R. and Anderson, D. A. and Powel, R. C. and Peterson, E. and Raithel, G.},
  month = dec,
  year = {2019},
  pages = {063427},
}

@article{lin_terahertz_2025,
  title = {Terahertz receiver based on room-temperature {Rydberg}-atoms},
  volume = {5},
  issn = {2667-3258},
  url = {https://www.sciencedirect.com/science/article/pii/S2667325823000705},
  doi = {10.1016/j.fmre.2023.02.019},
  number = {3},
  urldate = {2026-11-21},
  journal = {Fundamental Research},
  author = {Lin, Ya-Yi and She, Zhen-Yue and Chen, Zhi-Wen and Li, Xian-Zhe and Zhang, Cai-Xia and Liao, Kai-Yu and Zhang, Xin-Ding and Chen, Jie-Hua and Huang, Wei and Yan, Hui and Zhu, Shi-Liang},
  month = may,
  year = {2025},
  pages = {970--974},
}

@article{fancher_rydberg_2021,
  title = {Rydberg {Atom} {Electric} {Field} {Sensors} for {Communications} and {Sensing}},
  volume = {2},
  issn = {2689-1808},
  url = {https://ieeexplore.ieee.org/document/9374680},
  doi = {10.1109/TQE.2021.3065227},
  urldate = {2026-11-12},
  journal = {IEEE Transactions on Quantum Engineering},
  author = {Fancher, Charles T. and Scherer, David R. and John, Marc C. St. and Marlow, Bonnie L. Schmittberger},
  year = {2021},
  pages = {1--13},
}

@article{gordon_weak_2019,
  title = {Weak electric-field detection with sub-1 {Hz} resolution at radio frequencies using a {Rydberg} atom-based mixer},
  volume = {9},
  issn = {2158-3226},
  url = {https://pubs.aip.org/adv/article/9/4/045030/1076246/Weak-electric-field-detection-with-sub-1-Hz},
  doi = {10.1063/1.5095633},
  number = {4},
  urldate = {2024-09-11},
  journal = {AIP Advances},
  author = {Gordon, Joshua A. and Simons, Matthew T. and Haddab, Abdulaziz H. and Holloway, Christopher L.},
  month = apr,
  year = {2019},
  pages = {045030},
}

@inproceedings{hassun_effective_1997,
  title = {Effective evaluation of link quality using error vector magnitude techniques},
  url = {https://ieeexplore.ieee.org/document/622254},
  doi = {10.1109/WCC.1997.622254},
  urldate = {2026-11-23},
  booktitle = {Proceedings of 1997 {Wireless} {Communications} {Conference}},
  author = {Hassun, R. and Flaherty, M. and Matreci, R. and Taylor, M.},
  month = aug,
  year = {1997},
  pages = {89--94},
}

@misc{talashila_data_nodate,
  title = {Data for "{Communication} {Bandwidth} {Limits} of {Rydberg}-{Atom} {Receivers}"},
  url = {doi.org/10.18434/mds2-4413},
  urldate = {2026-11-24},
  author = {Talashila, Rajavardhan},
}

\clearpage

\appendix

\setcounter{figure}{0}
\renewcommand{\thefigure}{S\arabic{figure}}
\renewcommand{\theHfigure}{S\arabic{figure}}

\setcounter{equation}{0}
\renewcommand{\theequation}{S\arabic{equation}}
\renewcommand{\theHequation}{S\arabic{equation}}

\setcounter{table}{0}
\renewcommand{\thetable}{S\arabic{table}}
\renewcommand{\theHtable}{S\arabic{table}}

\section{System Characterization}
The system used in this work is comprehensively characterized, and the corresponding results are presented here.
\begin{figure}
    \centering
    \includegraphics[width=\linewidth]{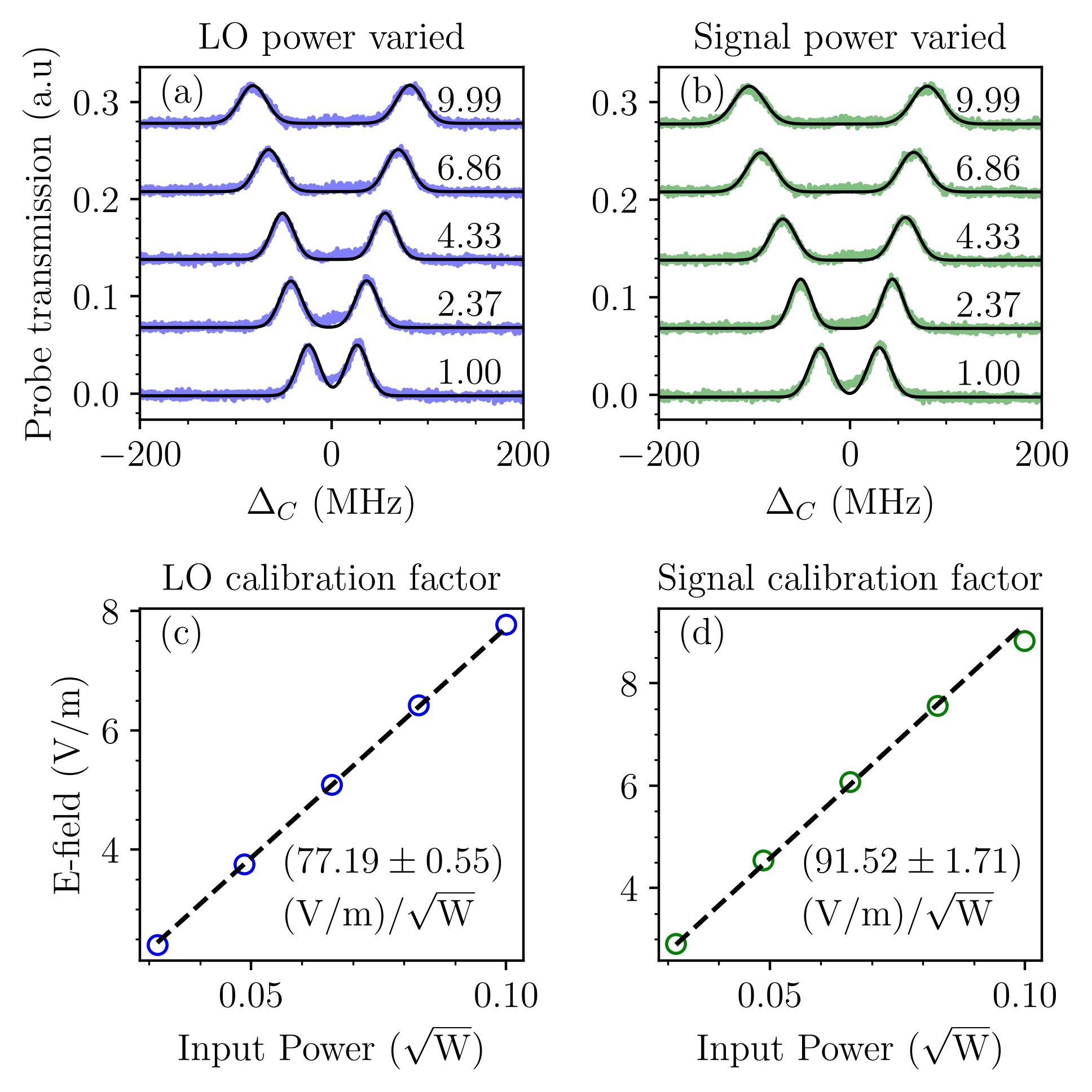}
    \caption{(a,b) The EIT spectra for various applied powers of  the $5709~MHz$ RF field generated by the LO and signal sources, respectively. The corresponding Gaussian fits are overlaid. The curves are offset for visual clarity. The input power in units of $\mathrm{mW}$ is indicated for each spectrum. (c,d) The calibration factors obtained by relating the applied RF power to the E-field inferred from the AT splitting of the EIT spectra for the LO and signal fields, respectively.  }
    \label{fig:Calibration}
\end{figure}

The RF field experienced by the atoms is calibrated using the AT splitting for both the signal and LO sources. In Fig. \ref{fig:Calibration}(a) and (b), the AT splitting curves obtained for various powers of the LO and signal are plotted, respectively. The electric field is determined from the experimentally observed AT splitting and calibrated against the applied input power, with the corresponding results shown in Fig. \ref{fig:Calibration}(c) and (d). The calculated calibration factors for the LO and signal are $(77.19\pm0.55)~(V/m)/\sqrt{W}$ and $91.52\pm1.71~(V/m)/\sqrt{W}$, respectively.

For the investigation of the  intermodulation properties, the probe and coupling lasers are frequency locked and RF frequency is fixed at an appropriate operating point. First, the optimum probe-laser detuning is determined. For this purpose, the coupling laser is swept over a broad range and the RF frequency is chosen such that the resulting beat-note spectral sidebands are symmetric. The beat note response is then measured systematically as a function of probe detuning, as shown in Fig.~\ref{fig:probe_couplin_RF_lock}(a). From these measurements, the probe laser is locked at the detuning corresponding to the maximum beat-note response, as shown in Fig.~\ref{fig:probe_couplin_RF_lock}(b).

Once the probe laser is locked, the RF frequency is varied and the corresponding beat-note response is recorded, as shown in Fig.~\ref{fig:probe_couplin_RF_lock}(c). The RF frequency and the coupling-laser detuning are then adjusted such that the beat-note response is symmetric about its center, and the coupling laser is locked at the center of this symmetric response. The resulting operating RF frequency is $5709~\mathrm{MHz}$.

\begin{figure}
    \centering
    \includegraphics[width=1\linewidth]{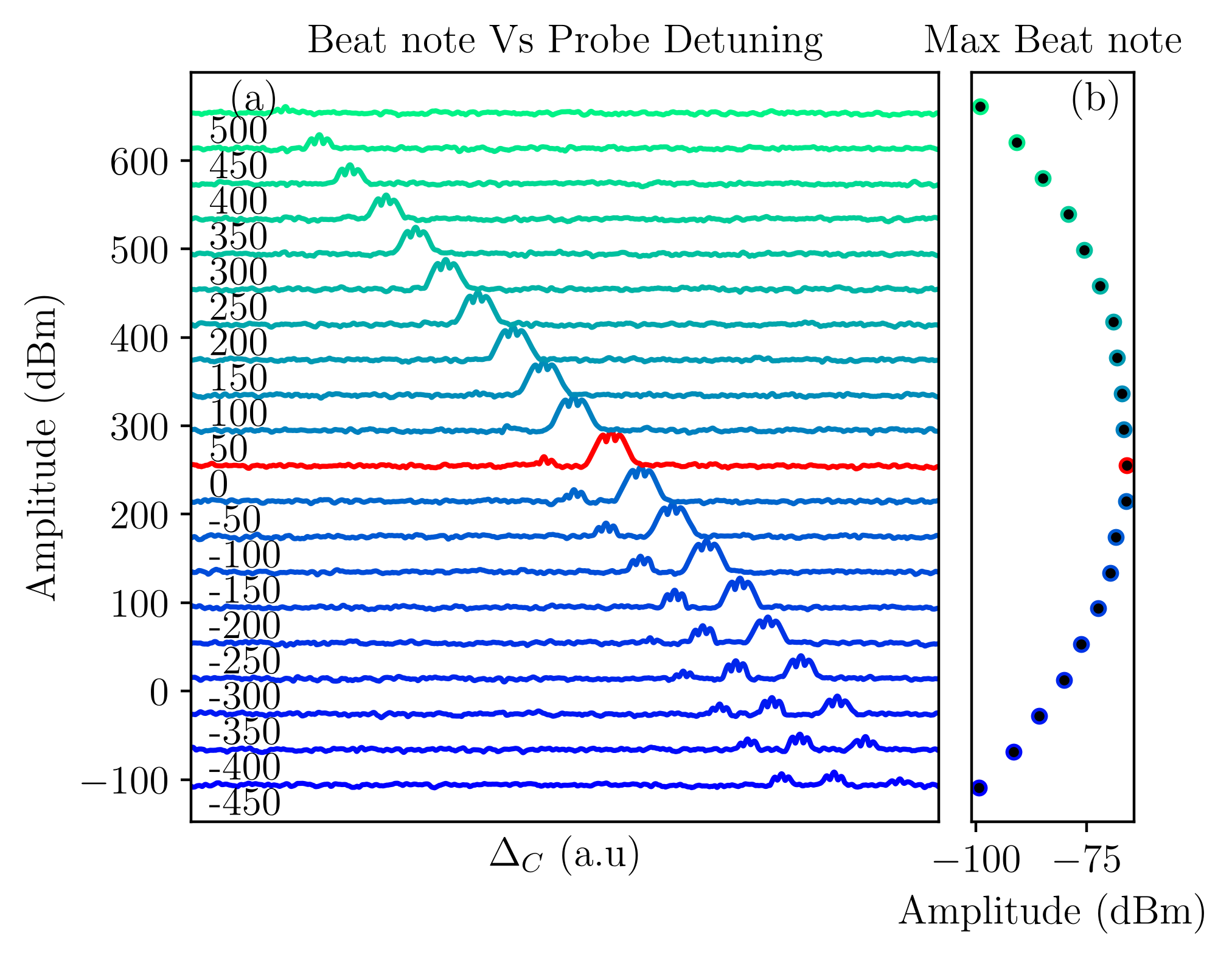}
    \includegraphics[width=1\linewidth]{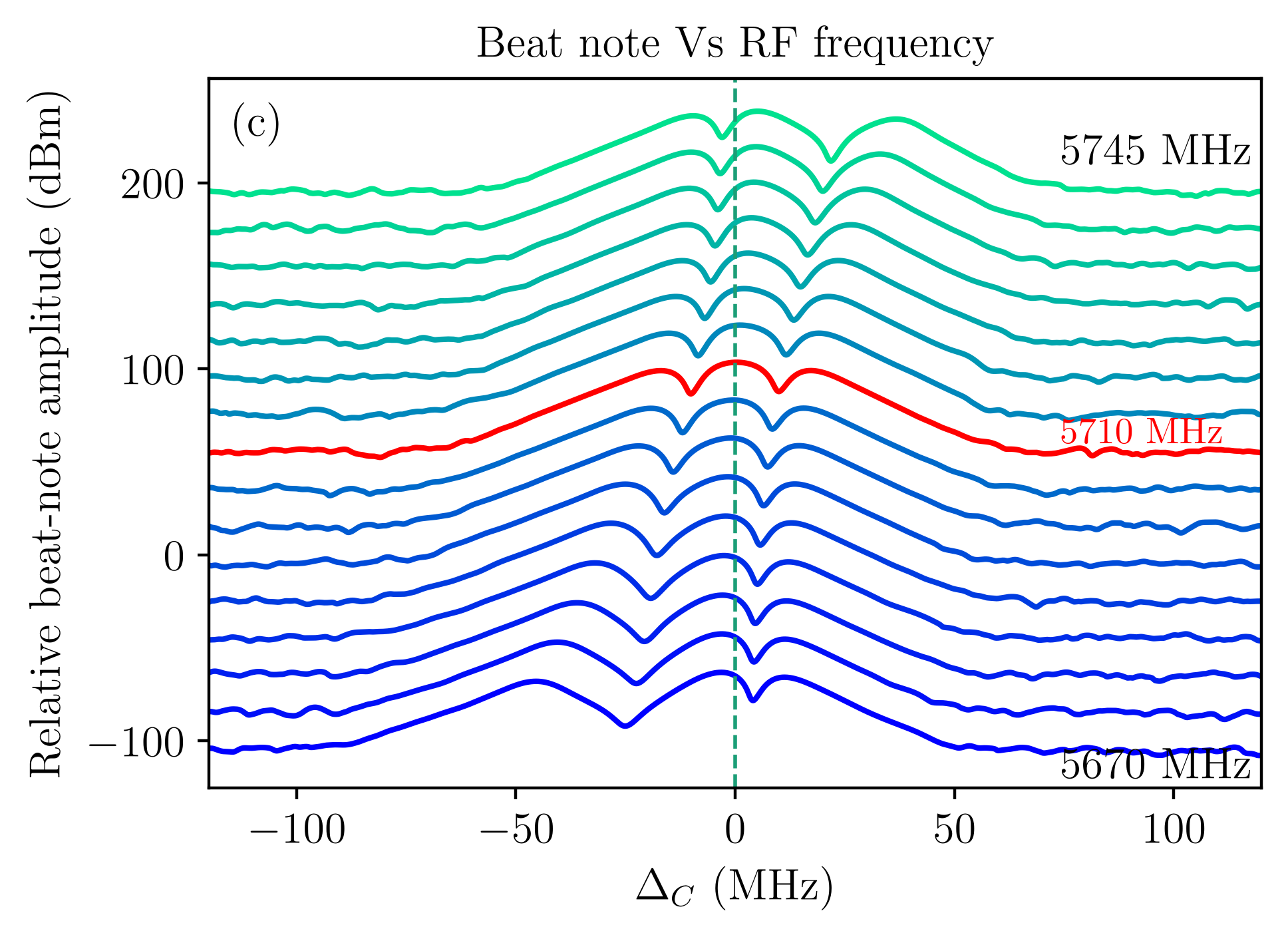}
    \caption{(a) Beat-note amplitude as a function of coupling-laser frequency detuning for different probe detunings. The traces are vertically offset by $40~\mathrm{dB}$ for visual clarity. (b) Maximum beat-note amplitude obtained at each probe-laser detuning. The probe laser is locked at the detuning corresponding to the maximum beat-note amplitude. (c) Determination of the optimum RF frequency. With the probe laser locked, the RF frequency is swept and the corresponding beat-note response is recorded. The optimum RF frequency is chosen as $5709~\mathrm{MHz}$ based on the symmetric beat-note response, and the coupling laser is subsequently locked at the center of this symmetric response for further measurements. }
    \label{fig:probe_couplin_RF_lock}
\end{figure}

\begin{figure}
    \centering
    \includegraphics[width=\linewidth]{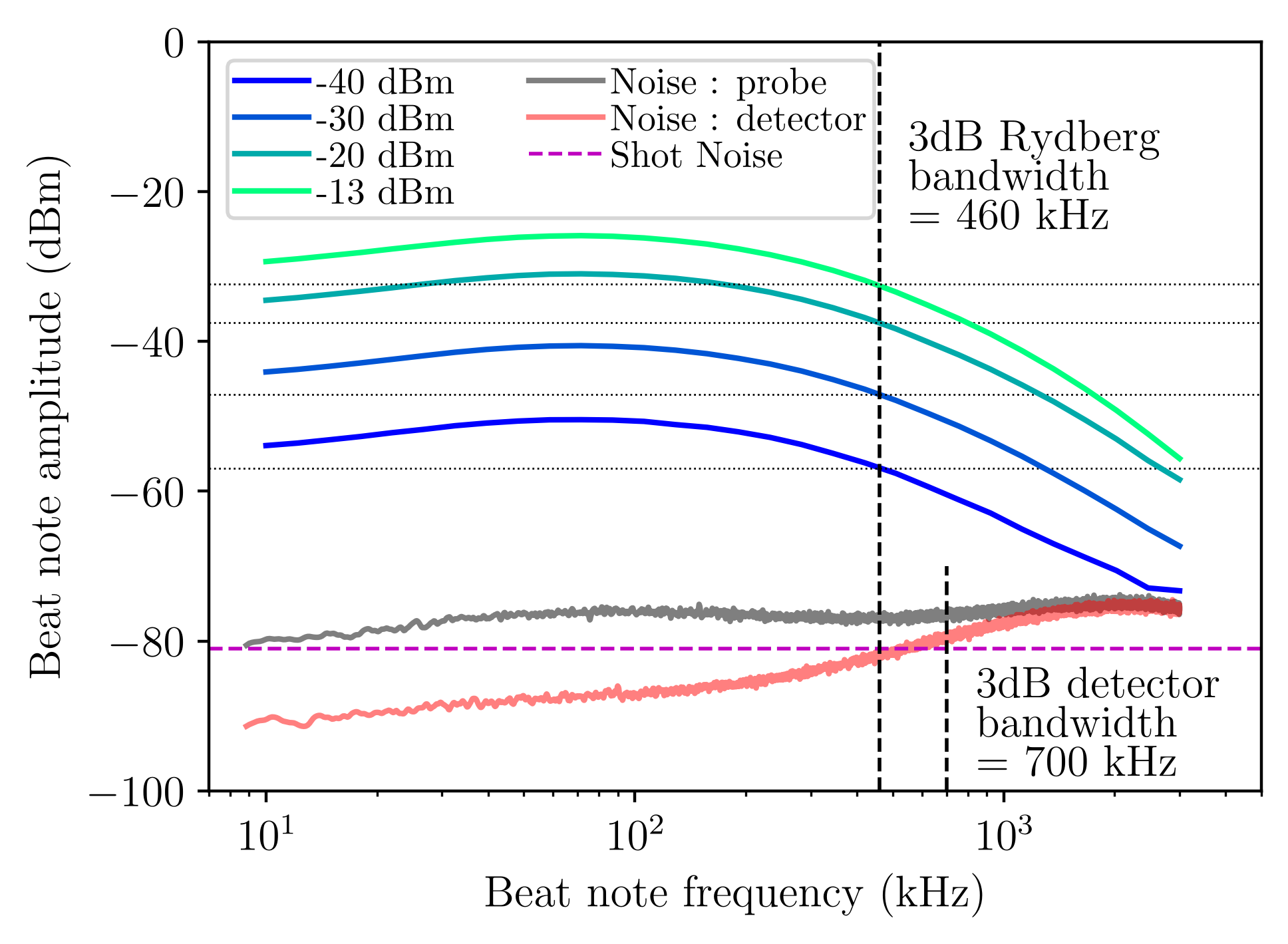}
    \caption{Characterization of the system bandwidth response. For different signal powers, the beat-note frequency is swept to determine the bandwidth of the Rydberg-atom receiver. The measurements yield a $3~\mathrm{dB}$ bandwidth of approximately $460~\mathrm{kHz}$. It is determined with respect to the beat-note power measured at the lowest investigated beat-note frequency. The noise floors measured with and without probe light incident on the photodetector are also shown. For the measurement with the probe light incident on the photodetector, the LO was applied at $-5~\mathrm{dBm}$ and signal field was turned off. For comparison, the $700~\mathrm{kHz}$,  $3~\mathrm{dB}$ bandwidth of the photodetector at the corresponding gain setting is indicated. The resolution bandwidth for all the measurements in this work is $1~\mathrm{kHz}$.}
    \label{fig:Bandwidth}
\end{figure}

The bandwidth of the Rydberg receiver for different signal powers is shown in Fig.~\ref{fig:Bandwidth}. The measurements confirm that the receiver bandwidth remains approximately unchanged even as the signal power approaches saturation. The measured receiver bandwidth is approximately $460~\mathrm{kHz}$ which is lower than the $700~\mathrm{kHz}$ bandwidth of the photodetector. The noise floors corresponding to the photodetector alone and to the probe-light induced noise are also plotted. The higher noise floors observed in the presence of the probe beam indicate that the system noise is dominated by the optical probe contribution rather than by the intrinsic electronic noise of the detection system.

For the shot noise calculations, the probe power falling on the photodetector is $\approx 11~\mathrm{\mu W}$.
Taking the responsivity of the photodetector at $850~\mathrm{nm}$ as $0.5~\mathrm{A/W}$, the photon current generated is $I_{\mathrm{ph}} = 5.5\mathrm{\mu A}$. Hence the shot-noise current spectral density is 
\begin{equation}
    i_{\mathrm{shot}} = \sqrt{2 q I_{\mathrm{ph}}} = 1.33~ \mathrm{pA/\sqrt{Hz}}.
\end{equation}

For the $1~\mathrm{kHz}$ resolution bandwidth used in the measurements of this work, the shot noise current is $\approx 42~\mathrm{pA}$.

For the photodetector used the transimpedance gain is $Z_T = 6.3\times 10^5 ~ \mathrm{V/A}$ corresponding to the generated noise voltage of 
\begin{equation}
    V_{\mathrm{shot,rms}} = 26~\mathrm{\mu V}.
\end{equation}

By taking into account the impedance considerations of the detector, the spectrum analyzer noise floor corresponding to the shot noise is 
\begin{equation}
    P_{\mathrm{shot}} = -81~\mathrm{dBm}.
\end{equation}
The calculated shot-noise floor is within a few decibels of the measured noise floor as shown in Fig.~\ref{fig:Bandwidth}.

A typical sensitivity measurement is shown in Fig.~\ref{fig:Sensitivity_Plots}(a). Similar measurements for the beat-note frequency range of 10 kHz to 600 kHz are presented in Fig.~\ref{fig:Sensitivity_Plots}(b).

\section{Beat note behavior derivation}
In line with the Gaussian approximation, the time variation of the EIT signal can be expressed as 
\begin{widetext}
\begin{align}
    B(\Delta_C,\Omega_S,\Omega_L,t) &= \frac{1}{2} \exp\left[ -\frac{1}{2\sigma^2}\left(\Delta_C - \frac{\Omega_L+\Omega_S \cos(\delta t + \phi)}{2}\right)^2\right] + \frac{1}{2} \exp\left[ -\frac{1}{2\sigma^2}\left(\Delta_C + \frac{\Omega_L+\Omega_S \cos(\delta t + \phi)}{2}\right)^2\right] \nonumber\\
    &=\bigg\{ \frac{1}{2} \exp\left[ -\frac{1}{2\sigma^2}\left(\Delta_C - \frac{\Omega_L}{2}\right)^2\right] \exp\left[\frac{(\Delta_C - \Omega_L/2)\Omega_S\cos(\delta t + \phi)}{2\sigma^2} \right] \nonumber \\
    &\qquad+ \frac{1}{2} \exp\left[ -\frac{1}{2\sigma^2}\left(\Delta_C + \frac{\Omega_L}{2}\right)^2\right] \exp\left[ -\frac{(\Delta_C + \Omega_L/2)\Omega_S\cos(\delta t + \phi)}{2\sigma^2} \right]  \bigg\} \exp\left[  -\frac{\Omega_S^2\cos^2(\delta t + \phi)}{8\sigma^2} \right].\nonumber
\end{align}

The factor at the end of this expression can be neglected to first order in $\Omega_S$. Additionally, we Taylor expand exponential factors with $\Omega_{S}$ in their arguments; while the full spectrum includes zeroth order terms from these expansions, the beat-note amplitude reduces to zero as $\Omega_{S}\rightarrow0$. Thus, we need only consider terms with (linear) $\Omega_{S}$ dependence, in which case,
\begin{align}
B(\Delta_C,\Omega_L,\Omega_S,t) &\simeq \frac{1}{2} \exp\left[ -\frac{1}{2\sigma^2}\left(\Delta_C - \frac{\Omega_L}{2}\right)^2\right] \left[ \frac{1}{2\sigma^2}\left(\Delta_C - \frac{\Omega_L}{2}\right)\Omega_S\cos(\delta t + \phi) \right]\nonumber \\
    &\qquad + \frac{1}{2} \exp\left[ -\frac{1}{2\sigma^2}\left(\Delta_C + \frac{\Omega_L}{2}\right)^2\right] \left[ -\frac{1}{2\sigma^2}\left(\Delta_C + \frac{\Omega_L}{2}\right)\Omega_S\cos(\delta t + \phi) \right] \nonumber \\
    &\simeq-\frac{1}{4}\left[ \frac{\partial \quad }{\partial\Delta_C} G_+(\Delta_C,\Omega_L) - \frac{\partial\quad}{\partial\Delta_C}  G_-(\Delta_C,\Omega_L)\right] \Omega_S \cos(\delta t + \phi).  \label{eq:EIT_Differentiation_Derivation}
\end{align}
The independent variable of the partial derivatives can be changed to $\Omega_L$, leading to the expression.
\begin{align}
B(\Delta_C,\Omega_L,\Omega_S,t) 
    &\simeq\frac{1}{2}\left[ \frac{\partial \quad }{\partial\Omega_L} G_+(\Delta_C,\Omega_L) + \frac{\partial\quad}{\partial\Omega_L}  G_-(\Delta_C,\Omega_L)\right] \Omega_S \cos(\delta t + \phi).  \label{eq:EIT_Differentiation_Derivation_B}
\end{align}

Both of the above expressions are mathematically equivalent. In the experimental implementation, the coupling laser is locked at a particular detuning and the applied field is varied to obtain the beat note signal.

\end{widetext}

\section{Optimum LO condition derivation}
Taking into consideration Eq.~\ref{eq:Beat_note_in_terms_of_Rabi_rate} at zero detuning we get
\begin{align}
    EIT_{AT}(\Delta_C=0,t) &= 
    \exp\left[ -\frac{\Omega_L^2+\Omega_S^2+2\Omega_S\Omega_L\cos\delta t}{8\sigma^2}\right].
\end{align}
Assuming $\Omega_S\ll\Omega_L$ and expanding the EIT response to first order in the signal amplitude, we get
\begin{align}
    B_0(t) &\simeq 
    -\exp\left[ -\frac{\Omega_L^2+\Omega_S^2}{8\sigma^2}\right] \left(\frac{\Omega_S\Omega_L\cos\delta t}{4\sigma^2}\right).
\end{align}

For finding the optimum LO Rabi rate $\Omega_L$ for maximum beat note amplitude, we take $\frac{\partial B_0(t)}{\partial \Omega_L}  = 0$, to obtain
\begin{align}
     \left( \frac{\Omega_S\cos\delta t}{4\sigma^2} - \frac{\Omega_S \Omega_L\cos\delta t }{4\sigma^2} \frac{2\Omega_L}{8\sigma^2} \right) &= 0  \nonumber\\
    \left( 1 - \frac{\Omega_L^2}{4 \sigma^2}\right) &= 0 \nonumber\\
    \Omega_L &= 2 \sigma.\nonumber
\end{align}
\begin{figure}
    \centering
    \includegraphics[width=\linewidth]{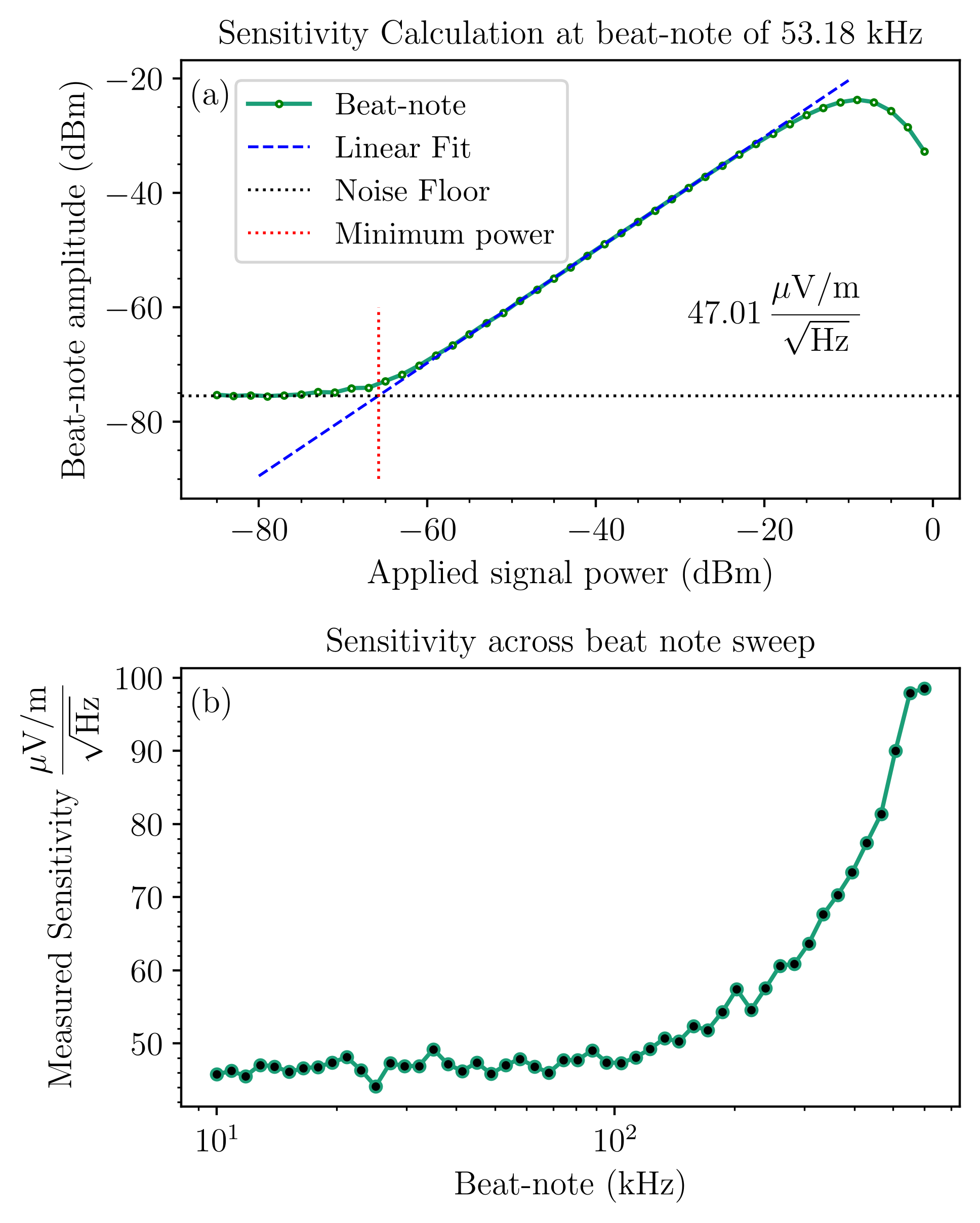}
    \caption{Sensitivity Plots. (a) The measurement of the sensitivity of the system. Here it is shown for $53.18~\mathrm{kHz}$. (b) The measured sensitivity by sweeping the beat-note.}
    \label{fig:Sensitivity_Plots}
\end{figure}

\section{Intermodulation Components Null conditions}
A representative figure showing the intermodulation components when a two-tone signal is fed to a nonlinear system is shown in Fig.~\ref{fig:IM_Representative}.

For two equal-amplitude signal tones, let the normalized LO and signal Rabi frequencies be
\begin{equation}
    L=\frac{\Omega_L}{\sigma},
    \qquad
    S=\frac{\Omega_s}{\sigma}.
\end{equation}
At zero coupling detuning, the nonlinear response Eq.~\ref{eq:Beat_Note_Nonlinear_zero_detuning} can be written as
\begin{align}
B(t)
&=
\exp\Bigg[
-\frac{1}{8}
\Big\{
L^2+2S^2
+2LS\left[\cos(\delta_1 t)+\cos(\delta_2 t)\right]
\nonumber\\
&\hspace{3.5cm}
+2S^2\cos\left[(\delta_1-\delta_2)t\right]
\Big\}
\Bigg].
\label{eq:two_tone_response}
\end{align}
Factoring out the LO-only contribution gives
\begin{equation}
B(t)
=
\exp\left
[-L^2/8\right]
\exp\left[
S A_1(t)+S^2 A_2(t)
\right],
\end{equation}
where
\begin{align}
A_1(t)
&=
-\frac{L}{4}
\left[
\cos(\delta_1 t)+\cos(\delta_2 t)
\right],\\
A_2(t)
&=
-\frac{1}{4}
\left[
1+\cos\left((\delta_1-\delta_2)t\right)
\right].
\end{align}

Any frequency component of interest, either fundamental component or 3rd order intermodulation component can be obtained from this nonlinear response equation from the corresponding complex Fourier coefficient. For example, the third order intermodulation product corresponding to $(2\delta_1-\delta_2)t$, can be obtained as
\begin{equation}
    C_{2,-1} = \frac{1}{(2\pi)^2}\int_0^{2\pi}\int_0^{2\pi} B(t) e^{-i(2\phi_1-\phi_2)} d\phi_1 d\phi_2,
    \label{eq:Fourier_3rd_Intermodulation}
\end{equation}
where $\phi_i = \delta_it$.
For simplifying the analytical expression, the exponential with the signal amplitude can be expanded to get 
\begin{align*}
\exp\left[S A_1+S^2A_2\right]
&= 
\sum_{p=0}^{\infty}
\frac{S^{p}}{p!}
A_1^p \times \sum_{q=0}^{\infty}
\frac{S^{2q}}{q!}
A_2^q \\
&= 
\sum_{p,q=0}^{\infty}
\frac{S^{p+2q}}{p!~q!}
A_1^p A_2^q.
\label{eq:S_expansion}
\end{align*}

Defining
\begin{equation*}
    u = e^{i\phi_1}, \qquad v = e^{i\phi_2},
\end{equation*}
we can write
\begin{equation*}
    A_1 = -\frac{L}{8}(u+u^{-1}+v+v^{-1}), \qquad A_2 = -\frac{1}{4} - \frac{1}{8}(uv^{-1}+vu^{-1}).
\end{equation*}

The third-order intermodulation component at
$2\delta_1-\delta_2$ is obtained by retaining only terms proportional
to $e^{i(2\phi_1-\phi_2)}$. For obtaining the coefficient $C_{2,-1}$, the integral in eq.~\ref{eq:Fourier_3rd_Intermodulation} keeps only terms of $e^{SA_1+S^2A_2}$ corresponding to $u^2v^{-1}$. As only odd orders of $S$ contribute to $(2\phi_1-\phi_2)$, we have the corresponding coefficients for $S^3$, $S^5$ and $S^7$ as
\begin{align*}
S^3 &: A_1A_2+\frac{A_1^3}{6}   \\
S^5 &: \frac{A_1 A_2^2}{2} + \frac{A_1^3A_2}{6} + \frac{A_1^5}{120}\\
S^7 &: \frac{A_1A_2^3}{6} + \frac{A_1^3A_2^2}{12} + \frac{A_1^5A_2}{120} + \frac{A_1^7}{5040},
\end{align*}

Considering only the $S^3$ terms, we get the coefficient
\begin{equation}
C_{2,-1}^{(3)}
=
\exp\left[-L^2/8\right]
\frac{L S^3}{1024}
\left(
16-L^2
\right).
\label{eq:IM3_leading}
\end{equation}
Thus, in the weak-signal limit, the nontrivial IM3 cancellation occurs at
\begin{equation}
L=4,
\qquad \implies
\Omega_{L,\mathrm{null}}=4\sigma.
\end{equation}

Retaining the higher orders $S^5$ and $S^7$, we get
\begin{align}
C_{2,-1}
\simeq
\exp&\left[-L^2/8\right] L S^3\nonumber\\
\times\Bigg[
&\frac{16-L^2}{1024} -
\frac{5S^2}{393216}
\left(
L^4-48L^2+384
\right)
\nonumber\\
&-
\frac{7S^4}{100663296}
\left(
L^6-96L^4+2304L^2-12288
\right)\nonumber\\
&+\mathcal{O}(S^6)
\Bigg].
\label{eq:IM3_finite_signal}
\end{align}

From this expression, the approximate condition for the null of third order intermodulation can be obtained by substituting 
\begin{equation}
L_{\mathrm{null}}^2
=
16+aS^2+bS^4+\mathcal{O}(S^6),
\end{equation}
and equating the coefficients  to Eq.~\eqref{eq:IM3_finite_signal} 
\begin{equation}
a=\frac{5}{3},
\qquad
b=\frac{1}{18}.
\end{equation}
Hence, the finite-signal IM3 cancellation condition is
\begin{equation}
\left(
\frac{\Omega_{L,\mathrm{IM3~null}}}{\sigma}
\right)^2
\simeq
16
+
\frac{5}{3}
\left(
\frac{\Omega_s}{\sigma}
\right)^2
+
\frac{1}{18}
\left(
\frac{\Omega_s}{\sigma}
\right)^4.
\label{eq:IM3_null_finite_signal}
\end{equation}

On similar lines, the finite-signal IM5 cancellation condition can be derived as
\begin{align}
\left(
\frac{\Omega_{L,\mathrm{IM5~null}}}{\sigma}
\right)^2
&\simeq
24 \pm 8\sqrt{3} + \frac{7}{12}(3\pm\sqrt{3})\left(
\frac{\Omega_s}{\sigma}
\right)^2 \nonumber\\
&\qquad+\frac{29}{5760}(6\pm\sqrt{3})
\left(
\frac{\Omega_s}{\sigma}
\right)^4.
\end{align}

On similar lines, the finite-signal IM7 cancellation condition can be derived as
\begin{align}
\left(
\frac{\Omega_{L,\mathrm{IM7~null}}}{\sigma}
\right)^2
&\simeq
7.48 + 0.42\left(
\frac{\Omega_s}{\sigma}
\right)^2+ 0.01\left(
\frac{\Omega_s}{\sigma}
\right)^4\\
\left(
\frac{\Omega_{L,\mathrm{IM7~null}}}{\sigma}
\right)^2
&\simeq
26.44 + 1.48\left(
\frac{\Omega_s}{\sigma}
\right)^2+ 0.026\left(
\frac{\Omega_s}{\sigma}
\right)^4\\
\left(
\frac{\Omega_{L,\mathrm{IM7~null}}}{\sigma}
\right)^2
&\simeq
62.07 + 3.49\left(
\frac{\Omega_s}{\sigma}
\right)^2+ 0.02\left(
\frac{\Omega_s}{\sigma}
\right)^4.
\end{align}

\begin{figure*}
    \includegraphics[width=\linewidth]{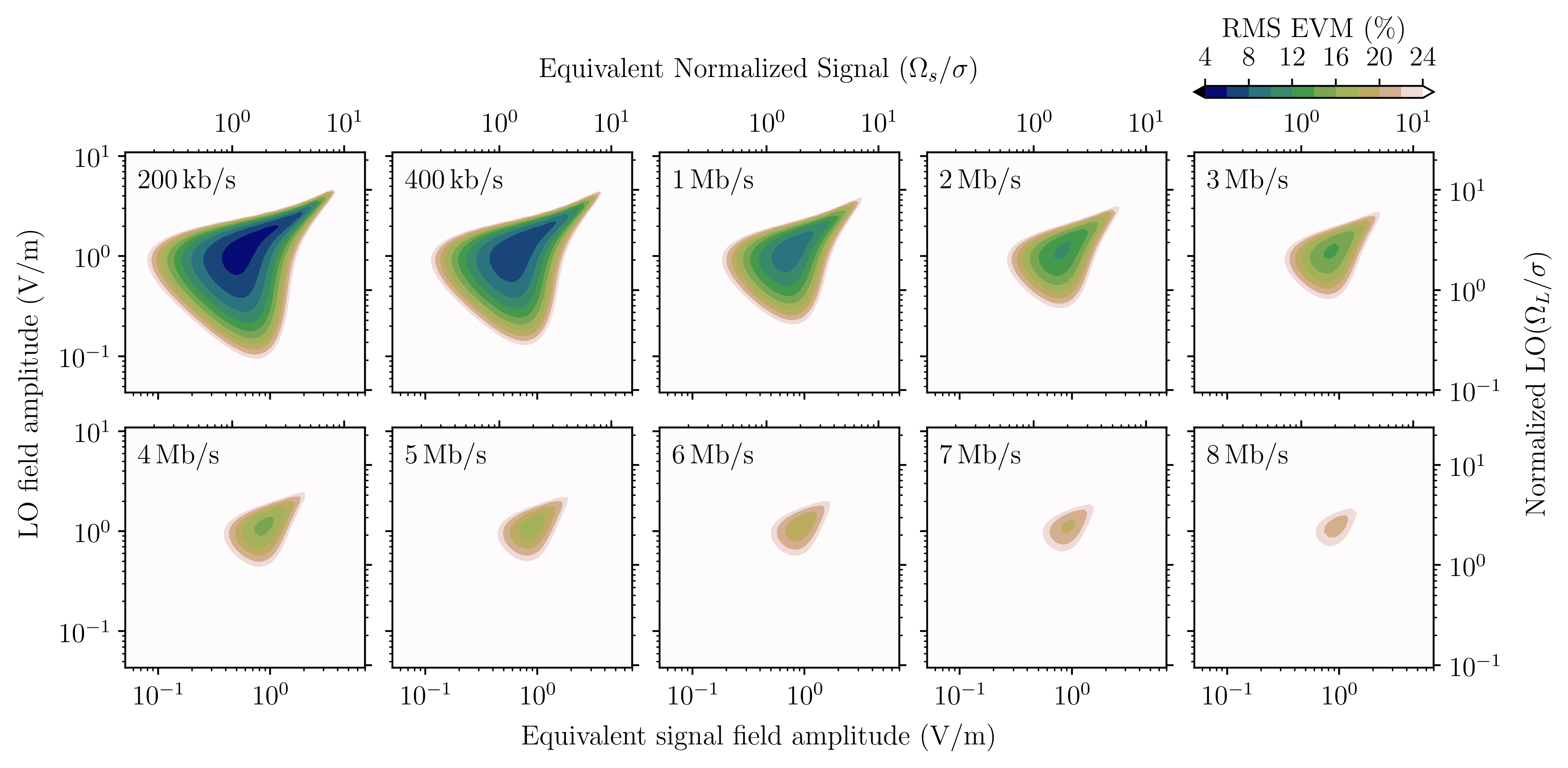}
    \caption{Modeled EVM for various transmitted bit rates. As the bit rate increases, achievable EVM degrades and also the usable range of signal and LO powers shrink.}
    \label{fig:EVM_bitrates_Summary}
\end{figure*}

\section{SNDR-EVM Relations}
In general, the signal-to-noise ratio (SNR) is defined as the ratio of the desired signal power $P_S$ to the noise power $P_N$ within the measurement bandwidth.

\begin{equation*}
    \mathrm{SNR} = \frac{P_S}{P_N}.
\end{equation*}

For the nonlinear response considered in this work, in-band distortion further degrades the received signal. Assuming that the noise arising from the photodetection process and the nonlinear distortion components contribute independently, an effective signal-to-noise-plus-distortion ratio (SNDR) can be expressed as

\begin{equation*}
    \mathrm{SNDR} = \frac{P_S}{P_N + P_D},
\end{equation*}
where $P_D$ represents the total in-band distortion power contributed by intermodulation products such as IM3, IM5, and IM7.

The root-mean-square error vector magnitude (EVM) represents the square root of the residual error power normalized to the desired signal power. Under the assumption that the photodetection noise and nonlinear distortion contributions are uncorrelated,

\begin{align*}
    \mathrm{EVM^2_{rms}} &= \frac{P_N + P_D}{P_S} = \frac{1}{\mathrm{SNDR}}, \\
    \mathrm{EVM^2_{rms}} &\simeq \frac{P_N}{P_S}+\frac{P_D}{P_S}.
\end{align*}
For white noise spectral density $N_0$ and an equivalent receiver noise bandwidth $B_n$,
\begin{equation*}
    P_N = N_0 B_n,
\end{equation*}
such that
\begin{equation*}
    \mathrm{EVM_{rms}^2} \simeq  \frac{N_0B_n}{P_S}+\frac{P_D}{P_S}.
\end{equation*}

\begin{figure*}[htbp]
    \centering
    \includegraphics[width=\textwidth]{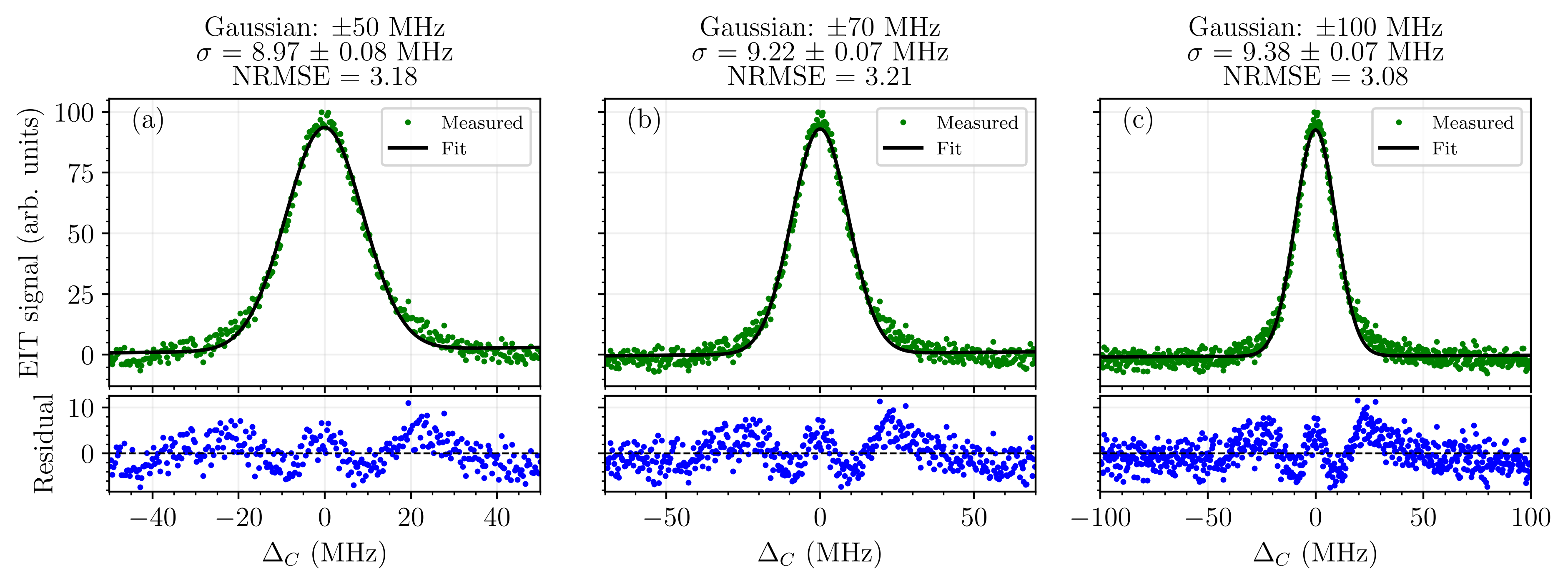}
    \caption{Comparison of Gaussian fits to the experimentally measured EIT
    spectrum over detuning ranges of $\pm 50$, $\pm 70$,
    and $\pm 100$ MHz. The measured data and fitted
    functions are shown along with the corresponding
    residuals. The fitting function includes a linear
    baseline. The fitted width parameters and NRMSE values
    are indicated in each panel.}
    \label{fig:Gaussian_Lorentzian}
\end{figure*}
For a specified maximum allowable EVM, $\mathrm{EVM_{req}}$, the maximum equivalent noise bandwidth is constrained by 

\begin{equation}
    B_n \le \left[\mathrm{EVM_{req}^2} - \frac{P_D}{P_S} \right] \frac{P_S}{N_0}.
\end{equation}
This expression explicitly shows the maximum achievable receiver bandwidth for a specified EVM requirement in the presence of the nonlinear distortion.
As the in-band distortion power $P_D$ increases, the maximum allowable receiver bandwidth decreases. If $P_D/P_S > \mathrm{EVM_{req}^2}$, the specified EVM cannot be achieved irrespective of the receiver bandwidth.

\section{Gaussian Fitting}
\label{app:Gaussian_Lorentzian}

The experimentally measured EIT lineshape was fitted using a Gaussian  function to examine the agreement with the measured spectra. The fitting function includes a linear baseline and is given by

\begin{align}
    G(\Delta_C) &= A \exp\left[-\frac{(\Delta_C-\mu)^2}{2\sigma^2}\right]
    + m\Delta_C + c, 
\end{align}   
where $A$ is the amplitude, $\mu$ is the center frequency,
$\sigma$ is the Gaussian standard deviation parameter, and $m$ and $c$ represent the
linear baseline.

The fitting was performed over three different detuning
ranges of $\pm 50~\mathrm{MHz}$, $\pm70~\mathrm{MHz}$, and $\pm 100~\mathrm{MHz}$. The
normalized root mean square error (NRMSE) was calculated as

\begin{equation}
    \mathrm{NRMSE} =
    \frac{100}{y_{\max}-y_{\min}}
    \sqrt{\frac{1}{N}
    \sum_{i=1}^{N}
    \left(y_i-y_{\mathrm{fit},i}\right)^2},
    \label{eq:NRMSE_fit}
\end{equation}
where $N$ is the number of data points within the fitting
range, $y_i$ is the measured signal, and $y_{\mathrm{fit},i}$
is the corresponding fitted value.

\begin{table}[htbp]
    \centering
    \caption{Comparison of Gaussian fits
    for different detuning ranges.}
    \label{tab:Gaussian_Lorentzian}
    \begin{tabular}{ccc}
        \hline
        Fitting range & $\sigma$ & Gaussian \\
        (MHz) & (MHz) & NRMSE (\%)  \\
        \hline
        $\pm 50$  & 8.97 & 3.18  \\
        $\pm 70$  & 9.22 & 3.21  \\
        $\pm 100$ & 9.38 & 3.08  \\
        \hline
    \end{tabular}
\end{table}

Figure~\ref{fig:Gaussian_Lorentzian} shows the measured EIT spectrum along with the Gaussian fits and their corresponding residuals. From Table~\ref{tab:Gaussian_Lorentzian}, the Gaussian function yields NRMSE values between 3.08\% and 3.21\%. 

The fitted Gaussian $\sigma$ changes from 8.97 MHz to 9.38 MHz when the fitting range is increased from
$\pm 50$ MHz to $\pm 100$ MHz, corresponding to a variation of approximately 4.6\%. The Gaussian NRMSE remains
approximately 3\% for all three fitting ranges. These results show that the Gaussian function provides a good approximation to the measured lineshape.
\clearpage

\end{document}